\documentclass[reqno]{amsart}
\usepackage[T1]{fontenc}
\usepackage[utf8]{inputenc}
\usepackage{timet}
\usepackage{amsmath,mathtools,empheq}
\usepackage{MnSymbol}
\usepackage{graphicx}
\usepackage{matlab-prettifier}
\usepackage{tikz}
\usetikzlibrary{arrows,3d,calc,intersections,decorations.pathmorphing}
\usetikzlibrary{decorations.markings,circuits.ee.IEC}

\usepackage[round]{natbib}
\usepackage[margin=1.5in]{geometry}

\usepackage[colorlinks,linkcolor=black,urlcolor=black,citecolor=black,anchorcolor=black]{hyperref}

\usepackage{scalerel,stackengine}
\stackMath
\newcommand\reallywidehat[1]{%
\savestack{\tmpbox}{\stretchto{%
  \scaleto{%
    \scalerel*[\widthof{\ensuremath{#1}}]{\kern.1pt\mathchar"0362\kern.1pt}%
    {\rule{0ex}{\textheight}}
  }{\textheight}%
}{2.4ex}}%
\stackon[-6.9pt]{#1}{\tmpbox}%
}

\DeclareMathOperator{\di}{d\!}

\DeclareMathOperator{\Si}{Si}
\newcommand{\bx}{\mathbf{x}}

\usepackage{booktabs,multirow}

\usepackage{bm}

\makeatletter
\newlength{\negph@wd}
\DeclareRobustCommand{\negphantom}[1]{%
  \ifmmode
    \mathpalette\negph@math{#1}%
  \else
    \negph@do{#1}%
  \fi
}
\newcommand{\negph@math}[2]{\negph@do{$\m@th#1#2$}}
\newcommand{\negph@do}[1]{%
  \settowidth{\negph@wd}{#1}%
  \hspace*{-\negph@wd}%
}
\makeatother

\makeatletter
\def\@setauthors{%
  \begingroup
  \def\thanks{\protect\thanks@warning}%
  \trivlist
  \centering\large \@topsep30\p@\relax
  \advance\@topsep by -\baselineskip
  \item\relax
  \author@andify\authors
  \def\\{\protect\linebreak}%
  \authors%
  \ifx\@empty\contribs
  \else
    ,\penalty-3 \space \@setcontribs
    \@closetoccontribs
  \fi
  \endtrivlist
  \endgroup
}
\def\@settitle{\begin{center}%
  \baselineskip14\p@\relax
    \normalfont\LARGE
  \@title
  \end{center}%
}
\makeatother

\begin{document}

\title[Modeling of dipping interfaces]{Finite-difference modeling of 2-D wave propagation in the vicinity of dipping interfaces: a comparison of anti-aliasing and equivalent medium approaches}
\author[E.F.M. Koene]{Erik F.M. Koene}
\author[J. Wittsten]{Jens Wittsten}
\author[J.O.A. Robertsson]{Johan O.A. Robertsson}

\address[E.F.M. Koene]{ETH Z\"urich, Institute of Geophysics, CH-8092 Z\"urich, Switzerland.}
\email{erik.koene@erdw.ethz.ch}
\address[J. Wittsten]{Centre for Mathematical Sciences, Lund University, Box 118, SE-221 00 Lund, Sweden, and Department of Engineering, University of Bor{\aa}s, SE-501 90 Bor{\aa}s, Sweden.}
\address[J.O.A. Robertsson]{ETH Z\"urich, Institute of Geophysics, CH-8092 Z\"urich, Switzerland.}

\begin{abstract}
Finite-difference (FD) modeling of seismic waves in the vicinity of dipping interfaces gives rise to artifacts. Examples are phase and amplitude errors, as well as staircase diffractions. Such errors can be reduced in two general ways. In the first approach, the interface can be anti-aliased (i.e., with an anti-aliased step-function, or a lowpass filter). Alternatively, the interface may be replaced with an equivalent medium (i.e., using Schoenberg \& Muir (SM) calculus or orthorhombic averaging). We test these strategies in acoustic, elastic isotropic, and elastic anisotropic settings. Computed FD solutions are compared to analytical solutions. We find that in acoustic media, anti-aliasing methods lead to the smallest errors. Conversely, in elastic media, the SM calculus provides the best accuracy. The downside of the SM calculus is that it requires an anisotropic FD solver even to model an interface between two isotropic materials. As a result, the computational cost increases compared to when using isotropic FD solvers. However, since coarser grid spacings can be used to represent the dipping interfaces, the two effects (an expensive FD solver on a coarser FD grid) equal out. Hence, the SM calculus can provide an efficient means to reduce errors, also in elastic isotropic media.
\end{abstract}

\maketitle


\section{Introduction}

The finite-difference (FD) method is a powerful numerical technique to simulate seismic waves. Within the FD method, derivatives in the wave equation are replaced by FD approximations of limited accuracy. The resulting differences between the continuous and discrete equations lead to errors. One such error stems from the difficulty of representing fine velocity models on coarse FD grids. Typically, a velocity model follows the geology of the subsurface. Such a geology can, for example, contain planar horizons that dip and curve arbitrarily. The FD grid, conversely, is only defined at regularly spaced discrete points. Since the grid size governs the computational cost of FD simulations, the question arises: how can a fine velocity model be accurately represented on coarse FD grids? Many different approaches to address this question have been discussed in the literature, and we refer to \cite{moczo2014} for a comprehensive historical overview and state-of-the-art review of these approaches. We may generally group the different methods into four classes:

\begin{enumerate}
    \item A first solution is to lay the FD nodes onto the velocity model and use the material properties that coincide with the FD nodes. That is, we sample the velocity model at a coarse resolution. This method is simple, but also inaccurate: (1) the exact location of an interface cannot be represented by the coarse model, and (2) the coarse sampling of an inclined interface generates staircase diffractions \citep[e.g.,][]{muir1992modeling}. Hence, this method cannot accurately reproduce the presence, angle, and location of interfaces.
    \item An alternative approach is to relax the restriction to regular FD grids. By deforming and stretching the FD grid (i.e., through some conformal mapping), the FD grid can be made to lie exactly parallel or perpendicular to an interface \citep[for an example see, e.g.,][]{fornberg1988pseudospectral}. To keep using the FD method, a coordinate transformation is used to map between a regular and a deformed grid. The downsides of this method are three-fold. First, the coordinate transformations increase the computational cost of the FD simulation. Second, generating an appropriately deformed FD grid is not a trivial problem. Third, using deformed grids is feasible only in the case of a sufficiently smooth interface geometry. Additionally, analyses in \cite{mittet2017} suggest that modeling errors occur regardless of whether or not the interface coincides with an FD node or not. This would imply that errors even exist for grid-aligned interfaces, deformed or not. These methods will thus not be pursued in this paper.
    \item A third solution is to average material parameters within FD cells into `equivalent media'. The idea is that a certain `sub-cell' resolution may be reached in FD modeling, i.e., that a single FD node can represent the seismic properties and geometry of several rock layers around the FD node. In a seminal paper, \cite{muir1992modeling} obtained great results in such an approach with the \citeauthor{schoenberg1989calculus} (SM; \citeyear{schoenberg1989calculus}) calculus. We note that the SM calculus itself is an extension of the medium averaging proposed by \cite{backus1962long}. The downside of SM calculus is that the average of two isotropic media is not isotropic itself, and FD modeling of such lower symmetry media is computationally expensive. \cite{moczo20023d} and \cite{moczo2014} therefore proposed alternative averaging methods, which keep the average of two isotropic media orthorhombic, such that standard low-cost \cite{virieux1986} FD grids remain applicable \citep{kristekorthorhombic2016}. Finally, \cite{capdeville20101} developed a more elaborate averaging method called `homogenization'. The method of homogenization is in principle similar to the ideas of \cite{backus1962long}: heterogeneities in the earth model, on a scale much smaller than the seismic wavelength, can be replaced with an appropriately smoothed medium without resulting in significantly different modeled seismic traces. The homogenization procedure requires the separation between large and fine length scales, which in itself depends on the (minimum) wavelength of waves in the medium. This latter property is not a part of the other effective media methods described to this point. We refer to the literature for details on the generation of the homogenization operator and its implementation \citep{capdeville20101,capdeville20102,capdeville2015fast,capdeville2018elastic}. Homogenization will not be pursued in this paper, because of its high upfront computational cost and the requirement for additional elastostatic solvers for the so-called `cell problem'. We remark that this does not mean that homogenization is not a viable method to address the interfaces problem. We merely leave out homogenization in order to focus on the quicker and cheaper methods to rapidly generate FD models.
    \item A fourth solution for the sampling of fine velocity models onto coarse FD grids was proposed by \cite{mittet2017}, who applied a low-pass filter to a fine compliance model before sampling it onto the coarse FD grid. A notable difference in comparison to the other methods discussed so far is that cells in the proximity of interfaces also become modified (due to the smoothing effect of the anti-aliasing procedure).
\end{enumerate}
The methods discussed above are all ad-hoc in nature. For example, the equivalent media methods are low-frequency approximations to Hooke's law. And the method of \cite{mittet2017} deals with the question of how to sample a fine step-function onto a coarse grid. The fact that they can be of use in FD simulations is not obvious at the outset. We speculate that this stems from the difficulty of studying interface errors in heterogeneous FD schemes mathematically. We recall that a \textit{heterogeneous FD scheme} indicates that we use the same FD update stencil in the entire model domain (where the material properties can vary from FD node to FD node), without explicitly enforcing boundary conditions across material interfaces. In such instances, FD solutions in heterogeneous media can have aliasing artifacts. We can illustrate this with the following simple example. Consider that in the wave equation, we multiply a propagating field (e.g., the pressure gradient) with a material field (e.g., with the reciprocal of the density). Now say that the propagating field is described by $P_0\sin(x)$ and that the material field is described by $\rho_0^{-1}\sin(x)$. Then their multiplication leads to $P_0\rho_0^{-1}\sin^2(x)=P_0\rho_0^{-1}(1-\cos(2x))/2$. Note how the wavenumber of the resulting field is double that of the two constituent fields. The interplay between the propagating and material fields can thus yield aliased solutions, even if the two fields individually are not aliased. The aliasing of solutions, as described here, leads to artifacts that are hard to study or control.

FD errors in heterogeneous media are thus hard to analyze mathematically. However, errors can alternatively be studied numerically. For example, \cite{vishnevsky2014numerical} numerically find that FD modeling of inclined interfaces limits errors to first-order accuracy only. That is, halving the grid spacing only halves the error. However, the actual size of the errors is omitted in their study. Here, we will analyze the accuracy of methods (i), (iii), and (iv) outlined above. Computed solutions will be compared to reference solutions. In particular, we will focus on a welded and inclined interface for three cases: (1) an acoustic-acoustic interface, (2) an elastic isotropic-elastic isotropic interface, and (3) an elastic anisotropic-elastic anisotropic interface. The inclined interface means that the results generalize to arbitrary geologies -- not just horizontally stratified media. Different combinations (e.g., acoustic-elastic interfaces, or elastic-vacuum boundaries) are not tested, due to an underlying assumption in the FD method which is that all propagating fields are continuous. This property (the so-called `welded interface') is, for example, violated at fluid-solid boundaries. Hence, modeling of hybrid physical systems requires extra physical and numerical considerations \citep[see, e.g.,][for ways to deal with this problem]{van2002finite,qu2020fluid}. In this paper, we do not investigate such complications. For completeness, we note that additional methods have been developed to deal with interfaces in the case of physics beyond anisotropic wave propagation. Examples are viscoelastic \citep[e.g.,][]{kristek2019visco} and poroelastic media \citep[e.g.,][]{moczo2018poro,gregor2020poro}, including discontinuities between poroelastic and elastic materials.

In the next section, we introduce the staggered grid and Lebedev grid methods for numerical modeling. In section 3, we first introduce a simple and fast 1-D anti-aliased step-function, and additionally describe the methods proposed by \cite{mittet2017}, \cite{muir1992modeling}, and \cite{kristekorthorhombic2016}. In section 4, the comparisons between the different methods will be carried out. Finally, in sections 5 and 6, we discuss the results and summarize our findings.

\section{The modeling method}
\subsection{The wave equations}
Assuming Einstein's summation convention, we write the (generally anisotropic) elastic wave equation for space $\bx$ and time $t$ as
\begin{empheq}[left=\empheqlbrace]{align}
    \rho(\bx)\frac{\partial v_i(\bx,t)}{\partial t} &= \frac{\partial \sigma_{ij}(\bx,t)}{\partial x_j} + f_i(\bx,t),\label{eq:velocity}\\
    \frac{\partial \sigma_{ij}(\bx,t)}{\partial t} &= c_{ijkl}(\bx)\frac{\partial v_k(\bx,t)}{\partial x_l}.\label{eq:stressdiff}
\end{empheq}
In these equations, $\rho$ is the density, $v_i$ a component of the particle velocity vector, $\sigma_{ij}$ a component of the stress tensor, $f_i$ a component of the force source density vector, and $c_{ijkl}$ is a component of the stiffness tensor. Additionally, we limit ourselves to 2-D wave propagation with $\bx=(x_1,x_3)=(x,z)$. The symmetries in Hooke's law then allow us to write eq. \eqref{eq:stressdiff} as
\begin{equation}\label{eq:stressstrainvoigt}
    \frac{\partial}{\partial t}\begin{pmatrix} \sigma_{xx}(\bx,t) \\ \sigma_{zz}(\bx,t) \\ \sigma_{xz}(\bx,t) \end{pmatrix} = \begin{pmatrix} C_{11}(\bx) & C_{13}(\bx) & C_{15}(\bx) \\ C_{13}(\bx) & C_{33}(\bx) & C_{35}(\bx) \\ C_{15}(\bx) & C_{35}(\bx) & C_{55}(\bx) \end{pmatrix}\begin{pmatrix} \frac{\partial v_x(\bx,t)}{\partial x} \\ \frac{\partial v_z(\bx,t)}{\partial z} \\ \frac{\partial v_z(\bx,t)}{\partial x} + \frac{\partial v_x(\bx,t)}{\partial z}  \end{pmatrix}.
\end{equation}
The elements $C_{IJ}$ are components of the stiffness {matrix}. We consider two successive simplifications of wave physics by simplifying the stiffness matrix in eq. \eqref{eq:stressstrainvoigt}:
\begin{enumerate}
    \item In an orthorhombic medium (for which the phase slowness surface is described by an ellipsoid whose principle axes are aligned with the Cartesian coordinate axes) the stiffness matrix takes on the form
\begin{equation}
    \begin{pmatrix} C_{11} & C_{13} & C_{15} \\ C_{13} & C_{33} & C_{35} \\ C_{15} & C_{35} & C_{55} \end{pmatrix}_\text{orthorhombic} = \begin{pmatrix} C_{11} & C_{13} & 0 \\ C_{13} & C_{33} & 0 \\ 0 & 0 & C_{55} \end{pmatrix}.
\end{equation}
An example of such a system is for example a transverse isotropic medium with a vertical axis of symmetry. A special case is that of an isotropic medium,
\begin{equation}
    \begin{pmatrix} C_{11} & C_{13} & C_{15} \\ C_{13} & C_{33} & C_{35} \\ C_{15} & C_{35} & C_{55} \end{pmatrix}_\text{isotropic} = \begin{pmatrix} \lambda+2\mu & \lambda & 0 \\ \lambda & \lambda+2\mu & 0 \\ 0 & 0 & \mu \end{pmatrix},
\end{equation}
with $\lambda$ and $\mu$ as the Lamé parameters.
    \item In an acoustic medium, the stiffness matrix is defined with just one parameter,
    \begin{equation}
    \begin{pmatrix} C_{11} & C_{13} & C_{15} \\ C_{13} & C_{33} & C_{35} \\ C_{15} & C_{35} & C_{55} \end{pmatrix}_\text{acoustic} = \begin{pmatrix} \frac{1}{\kappa} & \frac{1}{\kappa} & 0 \\ \frac{1}{\kappa} & \frac{1}{\kappa} & 0 \\ 0 & 0 & 0 \end{pmatrix},
\end{equation}
where $\kappa$ is the compliance. Typically, we take $\sigma_{1}=\sigma_{3}=-p$ to define the resulting wave equation in terms of the acoustic pressure $p$.
\end{enumerate}

\subsection{The FD approximations}
\begin{figure}
    \centering
    \includegraphics[width=0.98\textwidth]{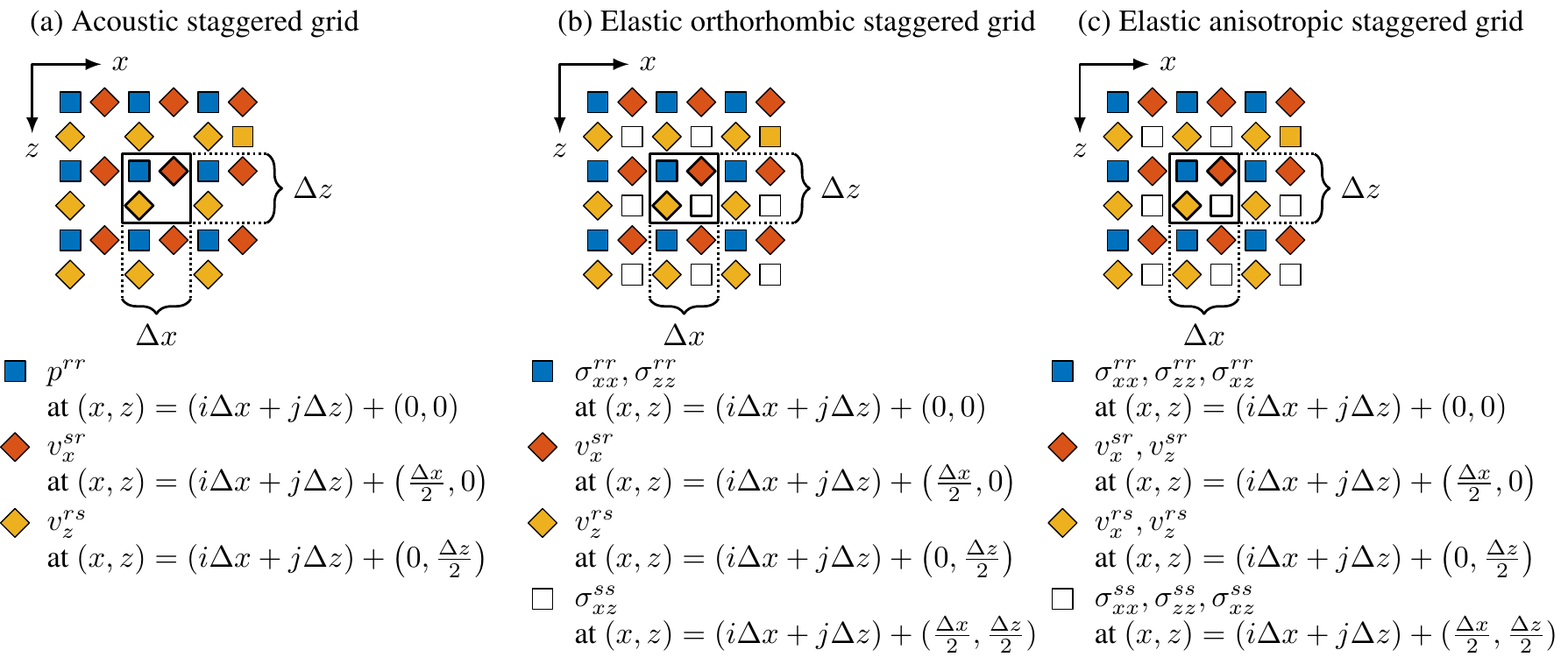}
    \caption{A schematic diagram showing the locations of stress (or pressure) and particle velocities in the three different staggered grids used in this paper. Note that the acoustic and orthorhombic layouts in panels (a) and (b) are essentially a subset of panel (c).}
    \label{fig:FDgrids}
\end{figure}
We will employ a staggered grid, which is the standard choice in FD modeling \citep{moczo2014}. We define staggered grid derivative operators $\text{D}_x$ and $\text{D}_z$ as
\begin{equation}\label{eq:stagderivative}
    \text{D}_x q(x,z) = \sum_{l=1}^L \alpha_l \frac{q(x\!+\!(l\!-\!\frac{1}{2})\Delta x,z)\!-\!q(x\!-\!(l\!-\!\frac{1}{2})\Delta x,z)}{\Delta x}
\end{equation}
and
\begin{equation}\label{eq:stagderivativeZ}
    \text{D}_z q(x,z) = \sum_{l=1}^L \alpha_l \frac{q(x,z\!+\!(l\!-\!\frac{1}{2})\Delta z)\!-\!q(x,z\!-\!(l\!-\!\frac{1}{2})\Delta z)}{\Delta z},
\end{equation}
such that $\text{D}_x$ approximates the horizontal partial derivative $\partial q/\partial x$ using FD coefficients $\alpha_l$ and a regular spacing of $\Delta x$. The operator $\text{D}_z$ works analogously. Quantities will be placed on the \cite{virieux1984sh,virieux1986} grid for acoustic and orthorhombic waves, and on the \cite{lebedev1964difference} grid for anisotropic waves \citep{lisitsa2010lebedev,bernth2011comparison}. The grids are displayed in Figure \ref{fig:FDgrids}. In the figure, and in the equations below, we employ additional superscripts to offset quantities following the staggered formalism. The double superscripts indicate the horizontal and vertical offsets, respectively. The superscript $^r$ (`reference') means no offset while the superscript $^s$ (`staggered') means an offset in the corresponding direction. For example, $v_x^{sr}(\bx,t)=v_x(x+\Delta x/2,z,t)$ is the horizontal particle velocity that is staggered in the horizontal direction but not in the vertical direction. Then we may write out the three fundamental modeling systems:
\begin{enumerate}
    \item For acoustic media, the staggered grid equations are
    \begin{empheq}[left=\empheqlbrace]{align}
    \rho^{sr}(\bx)\frac{\partial v_x^{sr}(\bx,t)}{\partial t} &\approx -\text{D}_x^{sr} p^{rr}(\bx,t) + f_x^{sr}(\bx,t),\\
    \rho^{rs}(\bx)\frac{\partial v_z^{rs}(\bx,t)}{\partial t} &\approx -\text{D}_z^{rs} p^{rr}(\bx,t) + f_z^{rs}(\bx,t),\\
    \kappa^{rr}(\bx)\frac{\partial p^{rr}(\bx,t)}{\partial t} &\approx -\text{D}_x^{rr} v_x^{sr}(\bx,t) -\text{D}_z^{rr} v_z^{rs}(\bx,t).
    \end{empheq}
    Note that following standard practice, we define the material properties (density $\rho$ and compliance $\kappa$) and the center of the FD operators ($\text{D}_x$ and $\text{D}_z$) at the same location as the quantity of which we take the temporal derivative. The layout of the grid is shown in Figure \ref{fig:FDgrids}a.
    \item For orthorhombic and isotropic media, the staggered grid equations are
    \begin{empheq}[left=\empheqlbrace]{align}
        \rho^{sr}(\bx)\frac{\partial v_x^{sr}(\bx,t)}{\partial t} &\approx \text{D}_x^{sr} \sigma_{xx}^{rr}(\bx,t) + \text{D}_z^{sr} \sigma_{xz}^{ss}(\bx,t) + f_x^{sr}(\bx,t),\\
        \rho^{rs}(\bx)\frac{\partial v_z^{rs}(\bx,t)}{\partial t} &\approx \text{D}_x^{rs} \sigma_{zz}^{rr}(\bx,t) + \text{D}_z^{rs} \sigma_{xz}^{ss}(\bx,t) + f_z^{rs}(\bx,t),\\
        \frac{\partial}{\partial t}\begin{pmatrix} \sigma_{xx}^{rr}(\bx,t) \\ \sigma_{zz}^{rr}(\bx,t) \\ \sigma_{xz}^{ss}(\bx,t) \end{pmatrix} &\approx \begin{pmatrix} C_{11}^{rr}(\bx) & C_{13}^{rr}(\bx) & 0 \\ C_{13}^{rr}(\bx) & C_{33}^{rr}(\bx) & 0 \\ 0 & 0 & C_{55}^{ss}(\bx) \end{pmatrix}\begin{pmatrix} \text{D}_x^{rr} v_x^{sr}(\bx,t) \\ \text{D}_z^{rr} v_z^{rs}(\bx,t) \\ \text{D}_z^{ss} v_x^{sr}(\bx,t) + \text{D}_x^{ss} v_z^{rs}(\bx,t) \end{pmatrix}.
    \end{empheq}
    The locations correspond to those shown in Figure \ref{fig:FDgrids}b.
    \item For generally anisotropic media, the staggered grid equations are
    \begin{empheq}[left=\empheqlbrace]{align}
        \rho^{sr}(\bx)\frac{\partial v_x^{sr}(\bx,t)}{\partial t} &\approx \text{D}_x^{sr} \sigma_{xx}^{rr}(\bx,t) + \text{D}_z^{sr} \sigma_{xz}^{ss}(\bx,t) + f_x^{sr}(\bx,t),\\
        \rho^{rs}(\bx)\frac{\partial v_x^{rs}(\bx,t)}{\partial t} &\approx \text{D}_x^{rs} \sigma_{xx}^{ss}(\bx,t) + \text{D}_z^{rs} \sigma_{xz}^{rr}(\bx,t) + f_x^{rs}(\bx,t),\\
        \rho^{rs}(\bx)\frac{\partial v_z^{rs}(\bx,t)}{\partial t} &\approx \text{D}_x^{rs} \sigma_{zz}^{rr}(\bx,t) + \text{D}_z^{rs} \sigma_{xz}^{ss}(\bx,t) + f_z^{rs}(\bx,t),\\
        \rho^{sr}(\bx)\frac{\partial v_z^{sr}(\bx,t)}{\partial t} &\approx \text{D}_x^{sr} \sigma_{zz}^{ss}(\bx,t) + \text{D}_z^{sr} \sigma_{xz}^{rr}(\bx,t) + f_z^{sr}(\bx,t),\\
        \frac{\partial}{\partial t}\begin{pmatrix} \sigma_{xx}^{rr}(\bx,t) \\ \sigma_{zz}^{rr}(\bx,t) \\ \sigma_{xz}^{rr}(\bx,t) \end{pmatrix} &\approx \begin{pmatrix} C_{11}^{rr}(\bx) & C_{13}^{rr}(\bx) & C_{15}^{rr}(\bx) \\ C_{13}^{rr}(\bx) & C_{33}^{rr}(\bx) & C_{35}^{rr}(\bx) \\ C_{15}^{rr}(\bx) & C_{35}^{rr}(\bx) & C_{55}^{rr}(\bx) \end{pmatrix}\begin{pmatrix} \text{D}_x^{rr} v_x^{sr}(\bx,t) \\ \text{D}_z^{rr} v_z^{rs}(\bx,t) \\ \text{D}_z^{rr} v_x^{rs}(\bx,t) + \text{D}_x^{rr} v_z^{sr}(\bx,t) \end{pmatrix},\\
        \frac{\partial}{\partial t}\begin{pmatrix} \sigma_{xx}^{ss}(\bx,t) \\ \sigma_{zz}^{ss}(\bx,t) \\ \sigma_{xz}^{ss}(\bx,t) \end{pmatrix} &\approx \begin{pmatrix} C_{11}^{ss}(\bx) & C_{13}^{ss}(\bx) & C_{15}^{ss}(\bx) \\ C_{13}^{ss}(\bx) & C_{33}^{ss}(\bx) & C_{35}^{ss}(\bx) \\ C_{15}^{ss}(\bx) & C_{35}^{ss}(\bx) & C_{55}^{ss}(\bx) \end{pmatrix}\begin{pmatrix} \text{D}_x^{ss} v_x^{rs}(\bx,t) \\ \text{D}_z^{ss} v_z^{rs}(\bx,t) \\ \text{D}_z^{ss} v_x^{sr}(\bx,t) + \text{D}_x^{ss} v_z^{rs}(\bx,t) \end{pmatrix}.
    \end{empheq}
    The locations correspond to those shown in Figure \ref{fig:FDgrids}c. This Lebedev grid modeling method thus requires nearly double the amount of memory and computational effort compared to the orthorhombic system. 
\end{enumerate}
In this paper, we will use a half-order $L=20$ with Taylor coefficients in eqs. \eqref{eq:stagderivative}--\eqref{eq:stagderivativeZ}. We opted for this large stencil size to ensure that no spatial dispersion errors are introduced by the modeling scheme, which could potentially confound the analysis of the interface errors. Hence, we want to isolate the errors due to the interface only. In practice, one can use smaller FD stencils to ensure a higher computational efficiency. Note that our findings are not necessarily biased by this large stencil size, as we will discuss later and show in Appendix \ref{app:smallerFDstencilresults}.

\subsection{Time-stepping and source injection}
A standard leap-frog time-stepping scheme will be used to step the solutions forward in time. Time-stepping errors are removed with time-dispersion transforms \citep{stork2013,mittet2017,mittet2019second,koene2018,wittsten2019}. In practice, that entails pre-processing the source-time function with the forward time-dispersion transform, and post-processing recorded traces with the inverse time-dispersion transform. The cost of these filters is negligible, and allows a use of large steps in time without loss of accuracy (provided that the simulation is stable, and that both the source and receiver traces are sampled adequately).

We employ the FD-consistent point-source as described in \cite{koene2020consistent} to excite point-sources on the FD grids. The FD-consistent point-source may be regarded as a wavenumber filter applied to a band-limited delta distribution. The filter is a function of the used FD coefficients. Its use corrects for amplitude-related errors that otherwise occur due to the injection of point-sources in FD grids. Note that for the Lebedev grid updates, it is implied that a source is injected into all four force-source components ($f_x^{rs},f_x^{sr},f_z^{rs},f_z^{sr}$) simultaneously, to avoid the creation of spurious waves \citep{lisitsa2011specific,koene2020lebedevsource}.

The described FD scheme is thus assumed to be highly accurate. Residual modeling errors are assumed to be due to the interface representations only. We will discuss the various interface representation methods in the next section.

\section{Interface representations}
\begin{figure}
    \centering
    \includegraphics[width=0.7\textwidth]{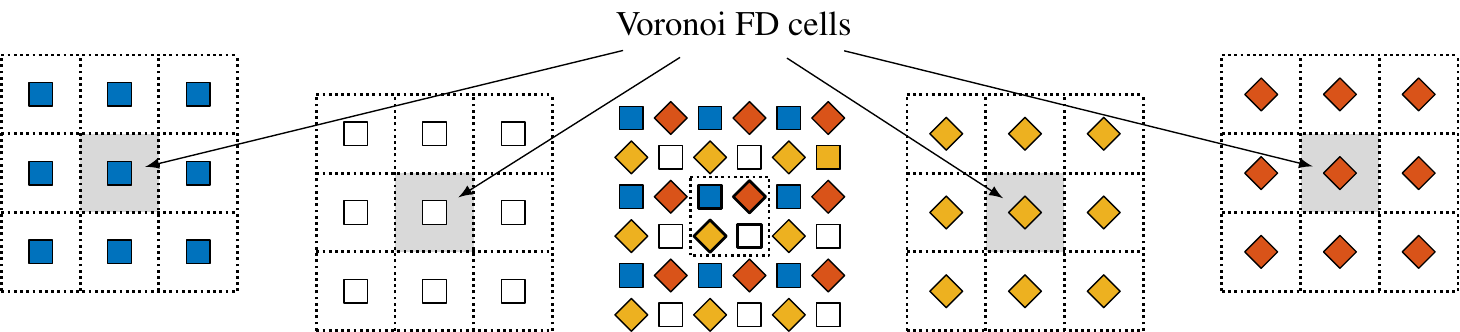}
    \caption{An illustration of the Voronoi FD cells. In the center, an elastic staggered FD grid is drawn (see Figure \ref{fig:FDgrids}). Four surrounding diagrams illustrate a Voronoi FD cell (shaded in gray), for each FD quantity individually.}
    \label{fig:voronoicell}
\end{figure}
\begin{figure}
    \centering
    \includegraphics[width=0.98\textwidth]{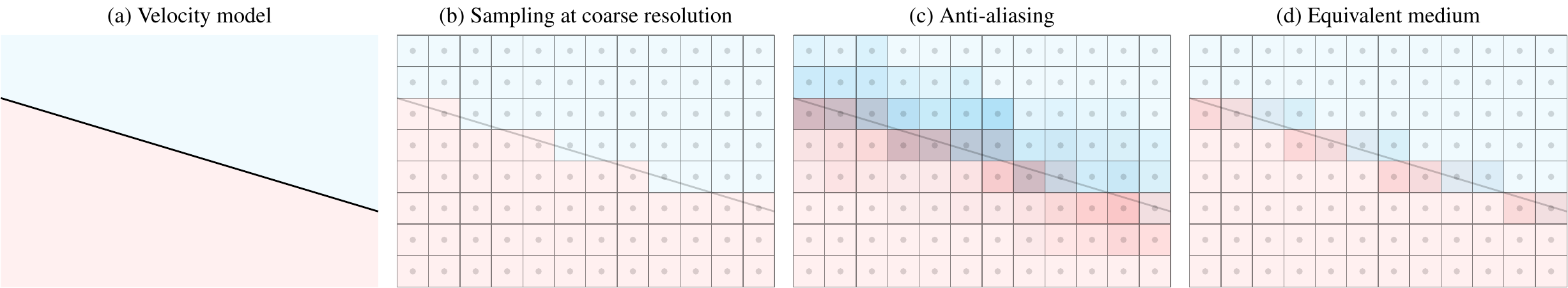}
    \caption{The set-up in this paper is an interface separating two media, shown in panel (a) through the use of two colors and a separating black line. Panel (b) displays the regularly spaced FD nodes (gray dots) inside the FD cells (gray squares). Material properties as present at the FD node are assigned to the entire FD cell. Panel (c) displays that we may smooth the sharp interface over multiple cells using an anti-aliasing filter. Panel (d) suggests that we may use an averaged `equivalent medium', only at cells intersected by the interface surface.}
    \label{fig:interfacessetup}
\end{figure}

We assume that each FD node lies at the center of a Voronoi cell. For example, Voronoi cells for the elastic staggered grid are illustrated in Figure \ref{fig:voronoicell}. Hence, although FD quantities are only defined at single points, we take them to be associated with a surrounding rectangular surface of size $\Delta x\times\Delta z$ in 2-D. Accordingly, we can fully partition a fine velocity model into a collection of cells. However, when modeling with the FD method, we must assign a \textit{constant} value to an entire FD cell. For example, consider a fine velocity model, as shown in Figure \ref{fig:interfacessetup}a. We now superimpose the FD nodes and FD cells onto the model. If we assign to each cell the material properties found at the FD node, we obtain Figure \ref{fig:interfacessetup}b, with a staircase appearance of the interface. An alternative procedure is shown in Figure \ref{fig:interfacessetup}c, where the interface is distributed over multiple nodes with an anti-aliasing procedure. Finally, Figure \ref{fig:interfacessetup}d shows that we may impose the presence of the interface only at intersected cells. Below, those methods are explained in more detail.

\subsection{Sampling at coarse resolution}
The simplest method to generate FD models is to sample the high-resolution velocity model at regular intervals. For example, in the acoustic case, we create a model by sampling the compliance model at regular intervals, i.e., $\kappa(i\Delta x,j\Delta z)$ for integers $i$ and $j$. Additionally, for horizontal particle velocity updates, we sample the density model as $\rho(i\Delta x+\Delta x/2,j\Delta z)$, to account for the staggered nature of the FD quantities. The density model for the vertical particle velocity model is analogously sampled at $\rho(i\Delta x,j\Delta z+\Delta z/2)$.

\subsection{Anti-aliased step-function}\label{sec:antialiasedstep}
\subsubsection{Theoretical considerations}
Consider that an interface in the fine model may be described with a 1-D Heaviside (or step-)function,
\begin{equation}\label{eq:heavisideinspace}
    H(z) = \begin{cases} 1 & \text{if }z>0 \\ \frac{1}{2}&\text{if }z=0, \\ 0 & \text{if }z<0, \end{cases}
\end{equation}
for example, a density model could take the form $\rho(x,z)=2000+1000H(z-1000)$. Now assume that we want to sample $H(z)$ onto the FD grid in a band-limited fashion, up to the Nyquist wavenumber $\pi/\Delta z$. \cite{mittet2017,mittet2018implementing} propose to compute such a step-function using fast Fourier transforms (FFTs). We, however, propose a method that is faster to compute and that is free of wrap-around artifacts. Following appendix \ref{app:antialiasedstepfunction}, the following expression corresponds to an anti-aliased step-function up to the Nyquist wavenumber $\pi/\Delta z$,
\begin{equation}\label{eq:heavisideSIfunction}
    H_\frac{\pi}{\Delta z}(z) = \frac{1}{2} + \frac{\Si\left( \frac{\pi}{\Delta z} z \right)}{\pi},
\end{equation}
where $\Si(z)=\int_0^z \sin(t)/t \di t$ is the sine integral. We can rapidly compute the $\Si(z)$ function using a rational polynomial, as detailed in Appendix \ref{app:antialiasedstepfunction}. We refer to equation \ref{eq:heavisideSIfunction} as the anti-aliased step-function.

\begin{figure}
    \centering
    \includegraphics[width=0.6\textwidth]{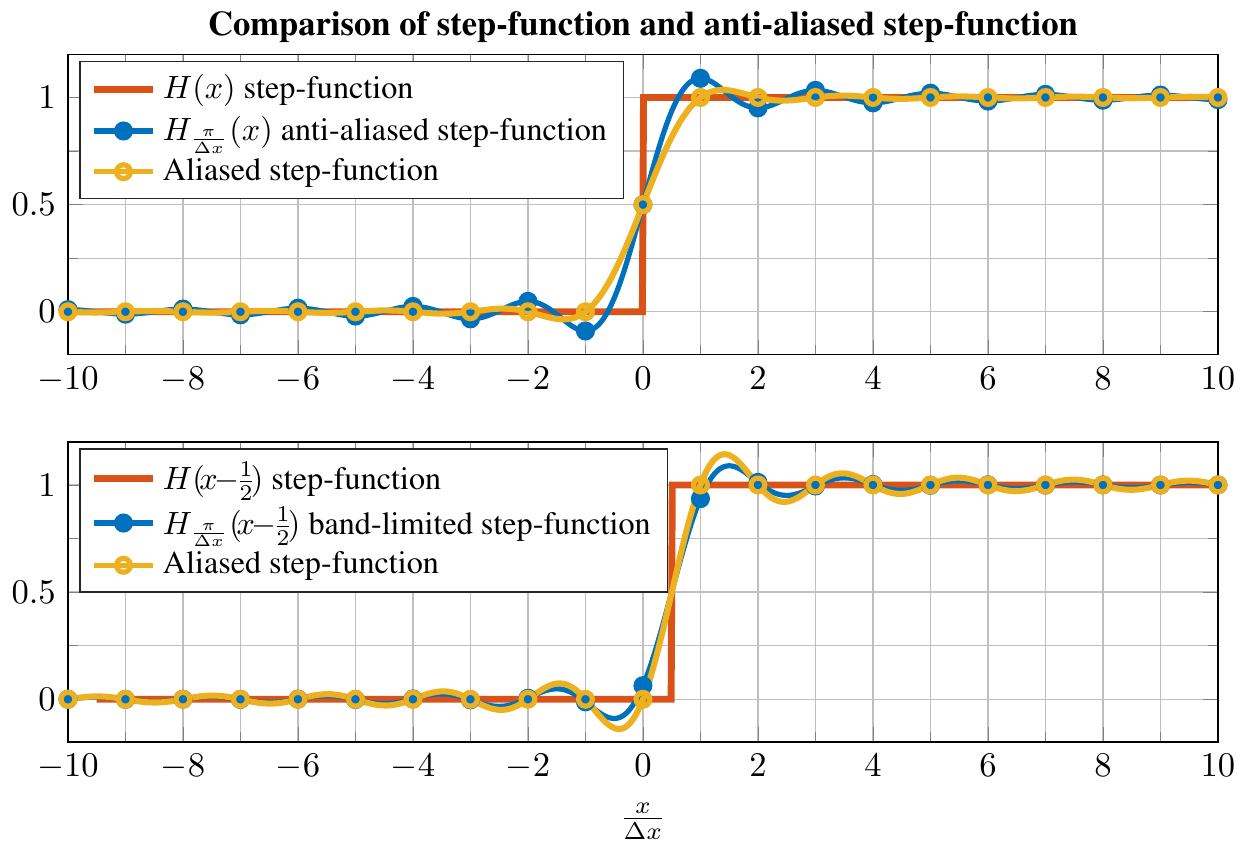}
    \caption{A plot that visualizes three versions of the Heaviside step-function: the standard step-function, an anti-aliased (i.e., band-limited) step-function with frequencies up to the Nyquist wavenumber $k_n=\pi/\Delta x$, and an aliased step-function that equals the standard step-function exactly on all integer steps $\Delta x$ and is interpolated with sinc interpolation elsewhere. The horizontal axis has been normalized to apply to arbitrary values for $\Delta x$.}
    \label{fig:bandlimitedheaviside}
\end{figure}

The anti-aliased step-function is plotted in Figure \ref{fig:bandlimitedheaviside}. Note that the anti-aliased step-function contains the Gibbs phenomenon: (1) overshoot at the step change and (2) ringing oscillations elsewhere. Additionally, we show a step-function that was sampled at integer nodes and then sinc-interpolated. The latter plot is referred to as the aliased step-function -- which is clearly different from its anti-aliased counterpart. \cite{mittet2017,mittet2018implementing} show that the use of anti-aliased step-functions leads to highly accurate results in 1-D FD wave simulations, while the standard Heaviside step-function leads to errors.

\begin{figure}
    \centering
    \includegraphics[width=0.5\textwidth]{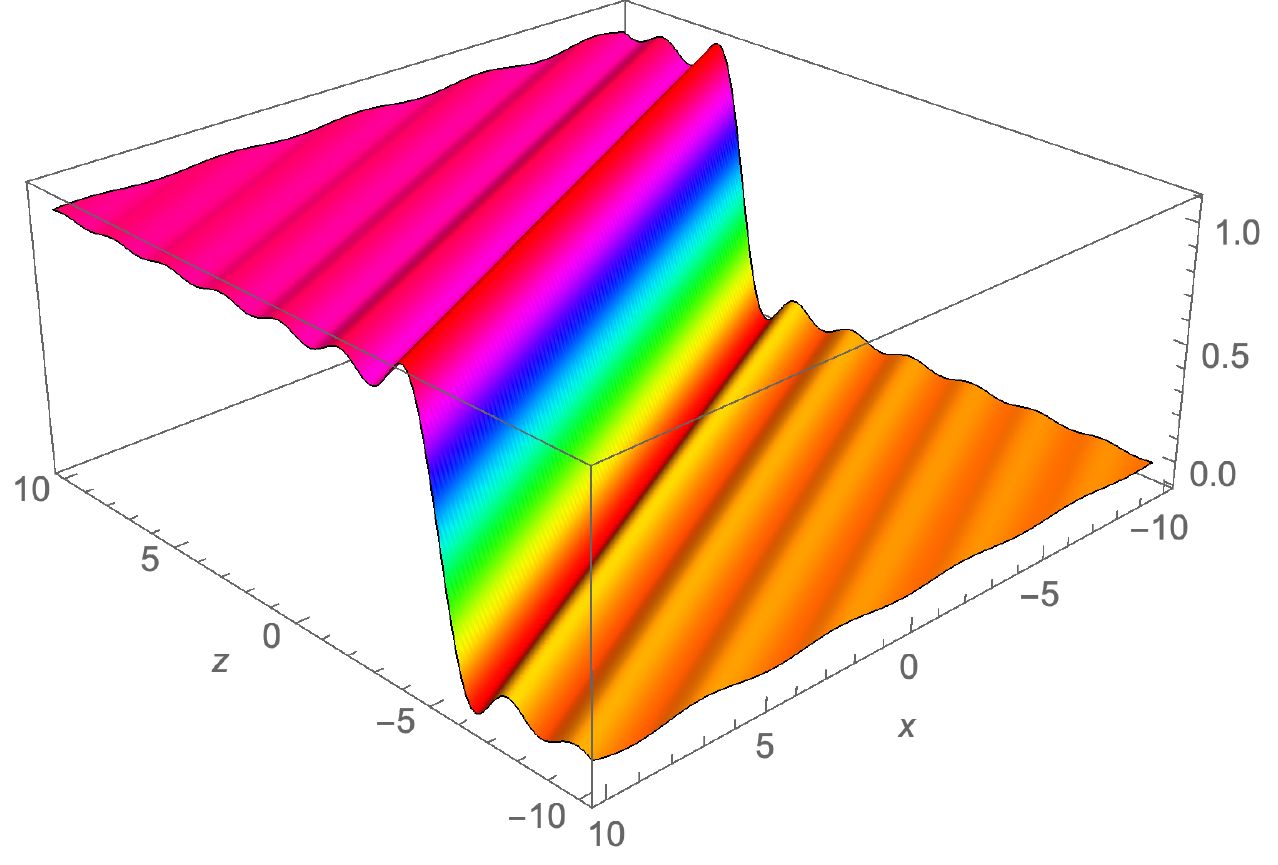}
    \caption{A (continuous) 3-D surface plot of a band-limited step-function $H_{\pi/1}([x+2z]/\sqrt{5})$, illustrating how a dipping interface may still be represented with a single band-limited step-function. For modeling, this function would be sampled at integer values.}
    \label{fig:3Dantialiasedstep}
\end{figure}

A constant inclined interface in 2-D can still be expressed with the use of a single step-function. One visualization of this principle is shown in Figure \ref{fig:3Dantialiasedstep}. All that is required is that we evaluate eq. \eqref{eq:heavisideSIfunction} with the distance to the interface. For example, an interface of $1\%$ is described by the equation of a line $z=x/100$. If we want to obtain the shortest signed (i.e., positive and negative valued) distance $d_s(x,z)$ from any point $(x,z)$ to this line, we can use simple algebra to find $d_s(x,z)=(-x+100 z)/\sqrt{100^2+1^2}$. Say that a density model is expressed as $\rho(x,z)=2000+1000H(d_s(x,z))$, then for FD modeling we simply replace the step-function with its anti-aliased counterpart, eq. \eqref{eq:heavisideSIfunction}. To account for potentially different grid spacings $\Delta x$ and $\Delta z$, we may use the normalized function $H_{\pi/1}$ evaluated at $d_s(x/\Delta x,z/\Delta z)$. Hence, the effective density model for this example would be written as $2000+1000H_{\pi}(d_s(x/\Delta x,z/\Delta z))$. This is the theoretical basis for the anti-aliased step-function.

\subsubsection{Practical implementation details}
In practice, we will use a window function to limit the extent of the ringing step-function. This is because the Gibbs phenomenon creates undesired oscillations also at nodes far away from the interface (as can be seen in Figure \ref{fig:bandlimitedheaviside} and Figure \ref{fig:3Dantialiasedstep}). Thus, the example developed in this section would be effectively computed using an additional window function $W$ as
\begin{align*}
    \rho_\text{anti-aliased step-function}(x,z) = \left(1-W\left[d_s\left(\frac{x}{\Delta x},\frac{z}{\Delta z}\right)\right]\right)\underbrace{\left[2000+1000H\left[d_s\left(\frac{x}{\Delta x},\frac{z}{\Delta z}\right)\right]\right]}_{\mathclap{\text{fine density model}}}& \\
    +W\left[d_s\left(\frac{x}{\Delta x},\frac{z}{\Delta z}\right)\right] \underbrace{\left[2000+1000H_{\pi}\left[d_s\left(\frac{x}{\Delta x},\frac{z}{\Delta z}\right)\right]\right]}_{\mathclap{\text{fine density model with anti-aliased step-function}}}&,
\end{align*}
where $d_s(x,z)$ again represents the signed distance between the interface and any point on the grid. In this study, we used a Kaiser window for $W$, with both the length and shape factor set to 3, but many other window functions will give similar results.

We have not yet specified which quantities to band-limit. Following \cite{mittet2017}, we will band-limit the density and the compliance matrix. The latter quantity may readily be obtained by taking the \textit{pseudo-}inverse ($C_{IJ}^+$) of the stiffness matrix $C_{IJ}$. We require the pseudo-inverse such that the acoustic compliance matrix may also be obtained. In appendix \ref{app:simplifiedcompliance} we provide explicit expressions for the compliance matrix in acoustic and elastic isotropic media.

We note that three issues may arise when modeling sharp boundaries with anti-aliasing:
\begin{enumerate}
    \item The method breaks down at fluid-solid interfaces. This may be easily seen from eq. \eqref{eq:isotropiccompliance}, where a $\mu^{-1}(\bx)$ grid is required for all locations, but with $\mu=0$ in the acoustic domain, we run into undefined behavior. We remark, however, that the FD method generally does not satisfy the boundary conditions for fluid-solid interfaces anyhow, such that other methods may be favored in such instances \citep[see, e.g.,][]{van2002finite}.
    \item At sharp boundaries, the sampled properties may become negative due to the overshoot in Gibbs phenomenon (e.g., a negative density may occur around an interface), which causes simulations to become unconditionally unstable. In such instances, one would be forced to clip the properties such that they are at least positive and not too small, such that they do not induce considerable spatial dispersion.
    \item As a counterpart, the sampled properties may end up with higher velocities than initially present, again due to the overshoot in Gibbs phenomenon. Due to the Courant-Friedrich-Lewy (CFL) condition, this may mean that the temporal sampling must be modified to ensure conditional stability of the FD simulation. This results in an increased computational cost.
\end{enumerate}
In the examples of this paper, we did not have to address any of these points, and merely point out the possibility of issues that may arise. There is no guidance from the literature yet concerning points (ii) and (iii), i.e., how to make sure that the anti-aliased medium does not result in instabilities or excessive dispersion.

\subsection{2-D anti-aliasing with lowpass filters \citep{mittet2017}}
The anti-aliased step-function does not easily lend itself for implementations in complicated media (e.g., in the case of curved interfaces). Hence, we may follow \cite{hobro2010rapid} and \cite{mittet2017} to note that we may instead obtain similar results using over-sampling followed by lowpass filtering of the model. Such an algorithm is straightforward to implement:
\begin{enumerate}
    \item Create a high-resolution oversampled grid of the density and compliance matrix for the entire model.
    \item Apply a zero-phase lowpass filter along each Cartesian dimension to the oversampled grids, to limit their wavenumber content around the desired Nyquist wavenumbers $\pi/\Delta x$ and $\pi/\Delta z$.
    \item Subsample the results at the final coarse resolution.
\end{enumerate}
Compared to the previous section, all practical advice remains essentially the same: (1) the use of a windowed lowpass filter can limit the spatial extent of the ringing, (2) we must fix potentially negative medium properties, (3) must watch out whether the CFL condition may be invalidated due to the filtering of the medium. The simplifications of eqs. \eqref{eq:acousticcompliancereinverted} and \eqref{eq:isotropiccompliancereinverted} are still applicable.

In the implementation used in this paper, we follow \cite{mittet2017} and oversample with a factor 10. The lowpass filter is applied as a finite impulse response (FIR) filter in a windowed approach. The filter is applied once along each Cartesian direction successively. A Hanning window of 51 points, centered around 1.1 times the desired Nyquist wavenumber, gave good results.

\subsection{Equivalent media with SM calculus \citep{muir1992modeling}}
The \citeauthor{schoenberg1989calculus} (SM; \citeyear{schoenberg1989calculus}) calculus assumes that the $z$ axis of the stiffness tensor lies perpendicular to the interface. For inclined interfaces, we must thus rotate the stiffness tensor to satisfy this requirement. After the equivalent medium is obtained, the stiffness matrix must then be rotated back to the original axes. Following \cite{schoenberg1989calculus}, we first partition the stiffness matrix into four parts,
\begin{align}
    \left(\begin{array}{c|cccc} 
    C_{11} & C_{13} &  C_{15} \\\hline 
    C_{13} & C_{33}  & C_{35} \\ 
    C_{15} & C_{35}  & C_{55} 
    \end{array}\right)_i = \left(\begin{array}{c|cccc} 
    C_{TT} & C_{TN} &   \\\hline 
    C_{TN}^T & C_{NN}  &  \\ 
      & &    &   
    \end{array}\right)_i,\label{eq:stiffnessfourmatrices}
\end{align}
for the $i$-th medium. We furthermore define the density $\rho_i$ for the $i$-th medium. The resulting equivalent medium is obtained as
\begin{align}
  \overline{\rho} &= \langle \rho \rangle, \label{eq:sm1} \\
  \overline{C_{NN}} &= \langle C_{NN}^{-1} \rangle^{-1}, \\
  \overline{C_{TN}} &= \langle C_{TN}C_{NN}^{-1} \rangle \overline{C_{NN}},\\
  \overline{C_{TT}} & = \langle C_{TT} \rangle - \langle C_{TN} C_{NN}^{-1} C_{TN}^T \rangle + \overline{C_{TN}}\langle C_{NN}^{-1} C_{TN}^T \rangle. \label{eq:sm4}
\end{align}
The angular brackets $\langle \dots \rangle$ indicate volumetric averaging. For example, if 40\% of an FD cell has density $\rho_0$ and the other 60\% of the cell has density $\rho_1$, then $\langle \rho \rangle=0.4\rho_0+0.6\rho_1$. 

After computing the averaged properties, the density of the equivalent medium becomes $\overline{\rho}$, and the submatrices $\overline{C_{NN}}$, $\overline{C_{TN}}$ and $\overline{C_{TT}}$ are back-substituted into eq. \eqref{eq:stiffnessfourmatrices} to obtain the equivalent stiffness matrix. In practice, the matrix inverses required in the SM calculus are best interpreted with the pseudoinverse, to allow for the averaging of acoustic data also. Note that the equivalent medium is transverse isotropic even when (unequal) `input' media are isotropic themselves. A tilted transverse isotropic stiffness matrix requires the expensive \cite{lebedev1964difference} grid modeling, even when the two isotropic media can individually be modeled on the \cite{virieux1986} grid.

\begin{figure}
    \centering
    \includegraphics[width=0.7\textwidth]{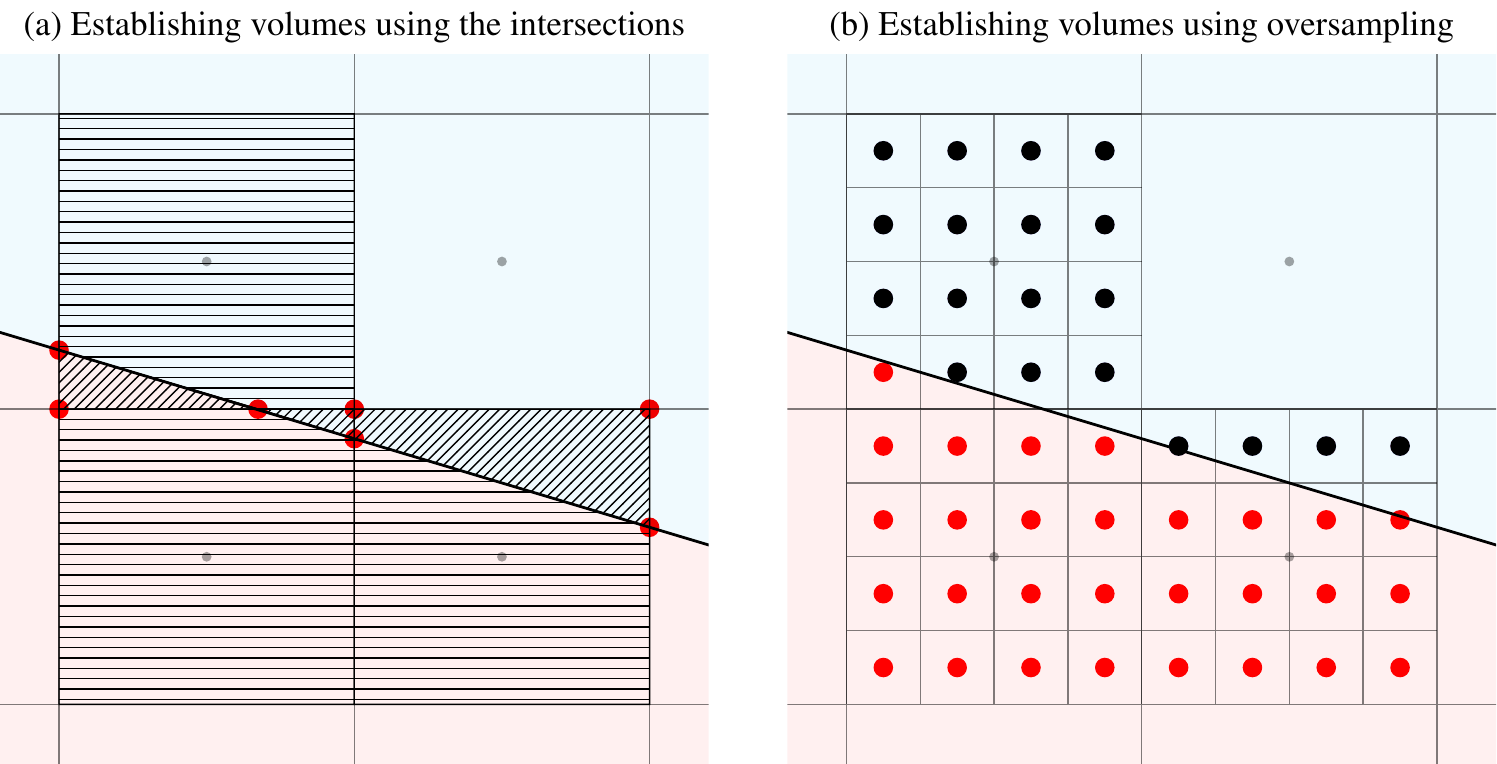}
    \caption{Zooming in on four FD cells. Panel (a) shows how the volumetric proportion of two media may be found from the geometry of the interface within the FD cells. Panel (b) illustrates how an oversampling procedure can be used to estimate the volumetric proportions of the intersected cells.}
    \label{fig:equivalentproportions}
\end{figure}

In summary, the algorithm for equivalent media with the SM calculus is:
\begin{enumerate}
    \item Generate a list of the cells containing a material interface. All other cells may be sampled at the coarse grid immediately.
    \item For each of the intersected cells, obtain an (estimate) of the volumetric proportions of medium 1 and medium 2 within it.
    \item If needed, rotate the stiffness tensors such that their (new) $z$ direction is perpendicular to the interface.
    \item Compute the equivalent medium following the SM calculus of eqs \eqref{eq:sm1}--\eqref{eq:sm4}.
    \item If needed, rotate the stiffness tensors back into the original coordinate system.
\end{enumerate}
In Figure \ref{fig:equivalentproportions}, two methods to estimate the volumetric proportions of media within an FD cell are shown. The method in Figure \ref{fig:equivalentproportions}a uses the geometry of the interfaces to compute these proportions. That is, the interface splits the FD cell into rectangles, triangles, and parallelograms. The areas of these shapes are easily computed and linked to proportions within cells. The method in Figure \ref{fig:equivalentproportions}b uses oversampling of the FD cells to approximate the volumes. We used the former method in this paper; \cite{hobro2010rapid} proposes a similar technique for 3-D media.

\subsection{Equivalent media with orthorhombic averaging \citep{moczo2014,kristekorthorhombic2016} }
Unlike the SM calculus, the technique proposed by \cite{moczo2014} keeps the $C_{15}$ and $C_{13}$ elements of the stiffness matrix zero, such that it is always possible to model the medium with the \cite{virieux1986} grid. Hence, the technique is called orthorhombic averaging, reflecting the form of the averaged stiffness matrix. This method equals the SM calculus for acoustic media. It also equals the SM calculus for isotropic media if interfaces are aligned with the grid. But the two methods differ for isotropic media with an inclined interface. We remark that this method of \cite{moczo2014} and \cite{kristekorthorhombic2016} considers in general also smooth heterogeneities in-between two interfaces. The orthorhombic averaging does not generalize to anisotropic media, virtually by design. The orthorhombic averaged stiffness tensor is
\begin{equation}
    \begin{pmatrix} C_{11} & C_{13} & C_{15} \\ C_{13} & C_{33} & C_{35} \\ C_{15} & C_{35} & C_{55} \end{pmatrix}_\text{orthorhombic averaging} = \begin{pmatrix} A & B & 0 \\ B & C & 0 \\ 0 & 0 & D \end{pmatrix},
\end{equation}
with
\begin{align}\label{eq:orthorhombic1}
    A & = \frac{\Delta x}{\Delta z}\left(\strokedint_{x} \left(\frac{1}{ \frac{\left(\strokedint_z \frac{\lambda}{\lambda+2\mu}\di z\right)^2}{\strokedint_{z} \frac{1}{\lambda+2\mu} \di z} + \strokedint_{z}\left( \lambda+2\mu - \frac{\lambda^2}{\lambda+2\mu} \right)\di z }\right) \di x\right)^{-1},\\
    B & = \frac{\strokedint_x\strokedint_z \frac{\lambda}{\lambda+2\mu}\di x\di z}{ \strokedint_x\strokedint_z \frac{1}{\lambda+2\mu}\di x \di z },\\
    C & = \frac{\Delta z}{\Delta x}\left(\strokedint_{z} \left(\frac{1}{ \frac{\left(\strokedint_x \frac{\lambda}{\lambda+2\mu}\di x\right)^2}{\strokedint_{x} \frac{1}{\lambda+2\mu} \di x} + \strokedint_{x}\left( \lambda+2\mu - \frac{\lambda^2}{\lambda+2\mu} \right)\di x }\right) \di z\right)^{-1},\\
    D & = \Delta x\Delta z\left(\strokedint_x\strokedint_z \frac{1}{\mu} \di x \di z\right)^{-1},\label{eq:orthorhombic4}
\end{align}
where $\strokedint_x\di x$ implies a numerical integration in the $x$ direction \textit{over just a single FD cell}. An analogous meaning is ascribed to $\strokedint_z \di z$. The value of $\strokedint_{x} \di x$ is thus equal to $\Delta x$ and $\strokedint_z \di z=\Delta z$. The density is similarly computed as $\rho=(\Delta x\Delta z)^{-1}\strokedint_x\strokedint_z \rho \di x\di z$ for each cell. The integrations are implemented with Riemann sums. 

In summary, the algorithm for an equivalent medium with the orthorhombic averaging procedure is:
\begin{enumerate}
    \item Generate a list of the cells containing an interface. All other cells may be sampled at the coarse grid immediately.
    \item Oversample the medium properties within each cell to numerically approximate the integrals in eqs. \eqref{eq:orthorhombic1}--\eqref{eq:orthorhombic4} with Riemann summations.
\end{enumerate}
In this paper, an oversampling factor of 8 was used in each Cartesian direction, following the suggestion of Jozef Kristek (personal communication, September 3, 2020). For reference, Figure \ref{fig:equivalentproportions}b uses an oversampling factor of 4 in each Cartesian direction. The authors of the orthorhombic averaging technique stress that their method is developed for a half-length $L=2$ FD scheme only, assuming at least 7 grid points per minimum wavelength. These two limits will not always be satisfied in this paper.

\subsection{Visual demonstration of the four techniques}
\begin{figure}
    \centering
    \includegraphics[width=0.85\textwidth]{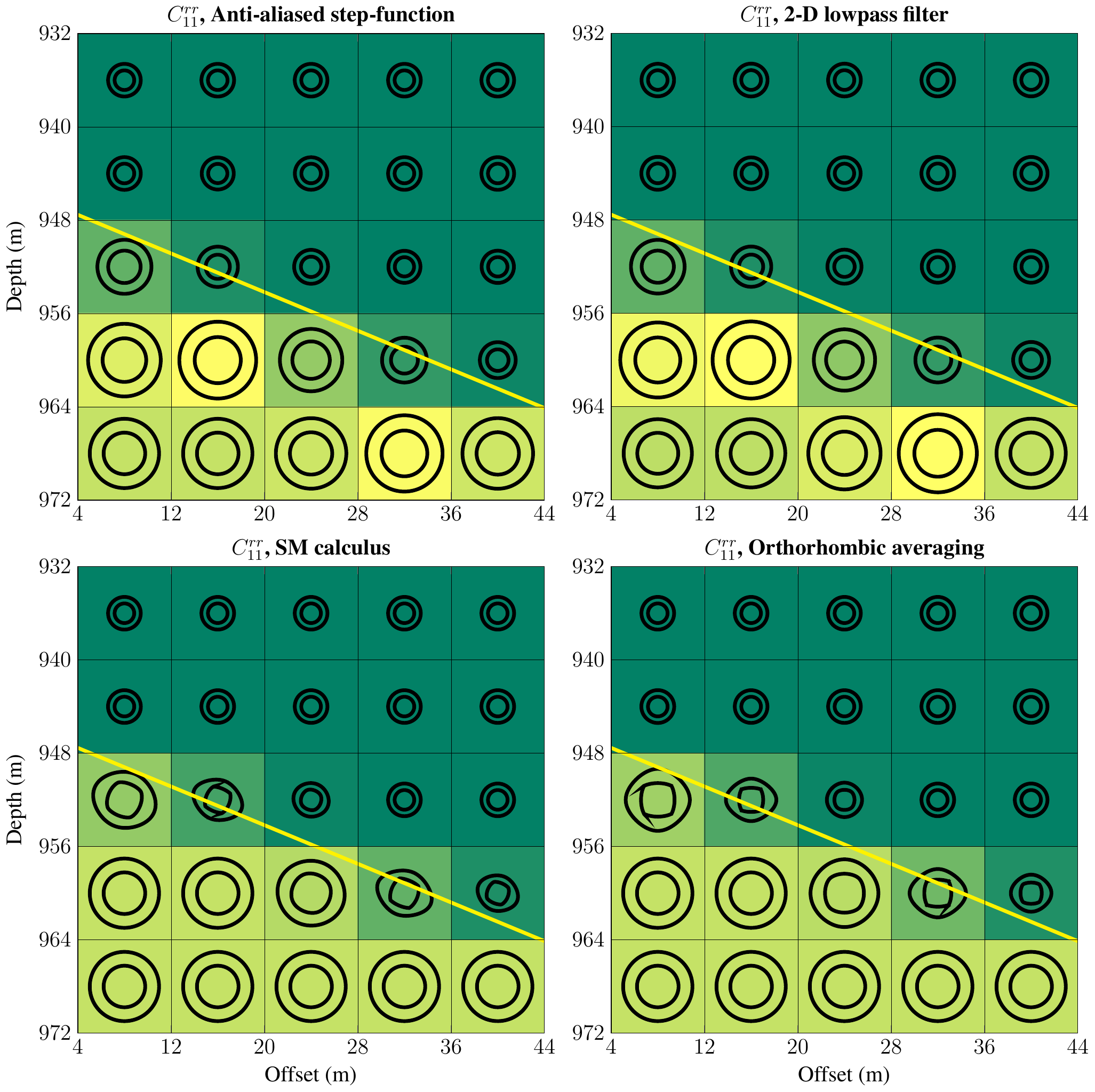}
    \caption{A demonstration of assigned $C_{11}^{rr}$ values (shown in color) by the four described techniques, in a $5\times 5$ grid of Voronoi FD cells with $\Delta x=\Delta z=8$ m, intersected by an interface (yellow line). Black curves in each cell correspond to the group velocity within each individual Voronoi FD cell (outer curves correspond to qP waves, inner curves to qS waves). Note how the anti-aliasing methods (top row) lead to fluctuations away from the interface, while the equivalent media methods (bottom row) lead to anisotropy for cells intersected by the interface.}
    \label{fig:effectivegroupvels}
\end{figure}

\begin{figure}
    \centering
    \includegraphics[width=0.85\textwidth]{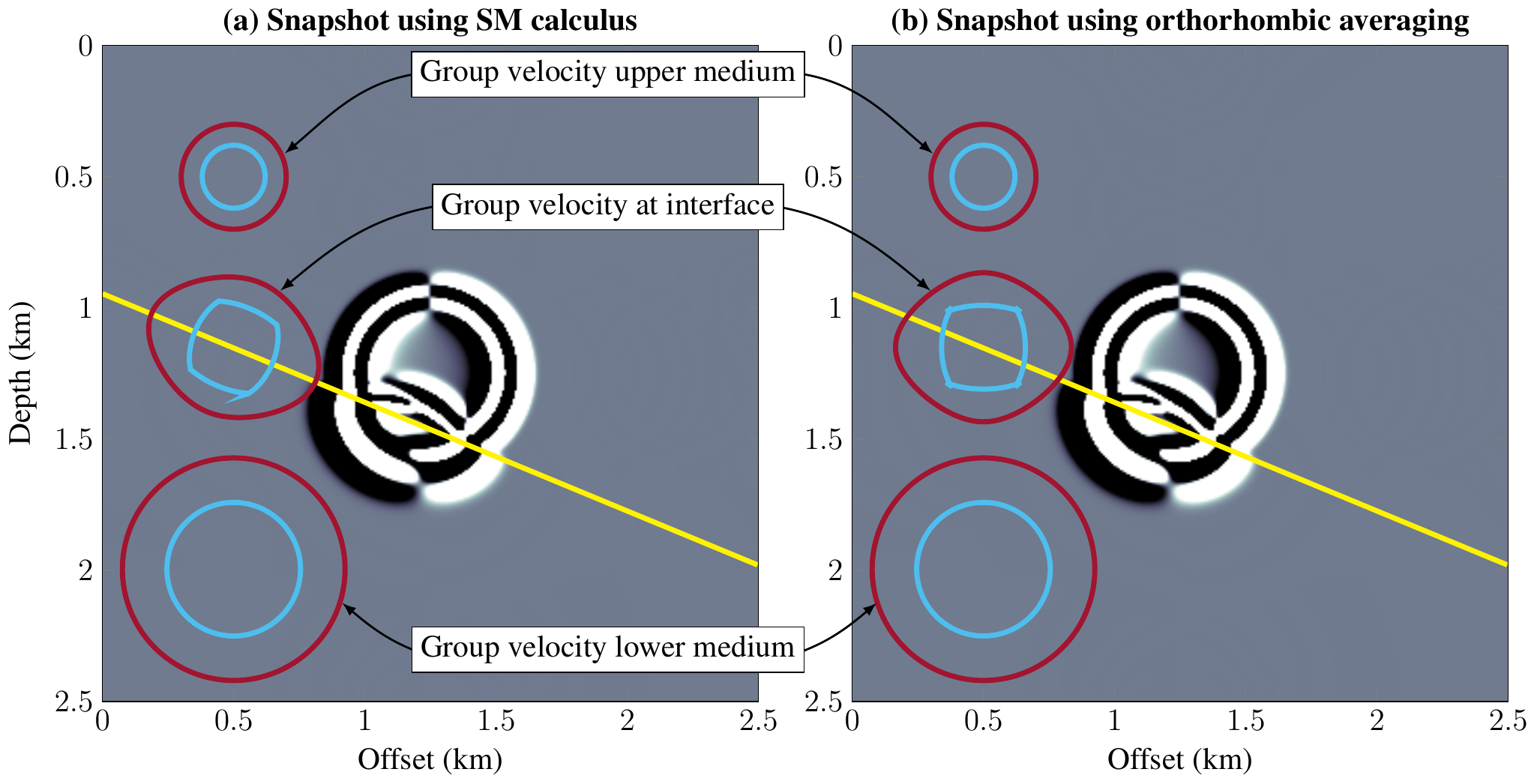}
    \caption{A demonstration of the difference between (a) the SM calculus and (b) the orthorhombic averaging. The arrows point to group velocity curves for the qP (red) and qS (blue) waves of the upper and lower medium, as well as the group velocity curves for a node at the interface between the two media. The SM calculus yields a tilted transverse anisotropic response at the interface, while the orthorhombic averaging yields an orthorhombic response.}
    \label{fig:contrast_SM_to_ortho}
\end{figure}

In Figure \ref{fig:effectivegroupvels} we provide a zoom of a velocity model partitioned with Voronoi FD cells, in the vicinity of an interface (compare to Figure \ref{fig:interfacessetup}). The upper region corresponds to a slow isotropic elastic medium, the lower region corresponds to a fast isotropic elastic medium. In the center of each cell, we illustrate the full assigned rock properties in the form of group velocity curves for qP and qS waves \citep[see, e.g.,][for a straightforward method to compute these curves in 2-D media]{carcione2012}. The anti-aliasing methods (in the top row of Figure \ref{fig:effectivegroupvels}) lead to fluctuations of the velocity model at cells near the interface. The fluctuations are larger below the interface (where $C_{11}^{rr}$ is larger, as it is the faster medium), which is related to the fact that we apply anti-aliasing on the inverse of the stiffness matrix, as shown in appendix \ref{app:simplifiedcompliance}. Conversely, the equivalent media methods (on the bottom row of Figure \ref{fig:effectivegroupvels}) show a modification of the velocity model only at cells intersected by the interface. The group velocity curves show how the properties assigned to these cells are anisotropic. In Figure \ref{fig:contrast_SM_to_ortho}, we similarly show how the SM calculus and orthorhombic averaging differ in the resulting group velocities of qP and qS waves at the interface: the SM calculus produces a tilted anisotropy aligned with the interface, while orthorhombic averaging creates anisotropy that is aligned with the modeling axes.

\section{Numerical tests}
We will now test the accuracy of the interface representations described above. For the sake of readability, we use a shorthand way of referring to the results. The \textit{interface treatment method} is synonymous to \textit{an FD simulation on an FD grid prepared with the interface treatment method}. For example, when writing ``the SM calculus results in small residual staircase diffractions'', it is implied that we mean that ``the wavefield that was computed with the FD method, after sampling the fine velocity model onto the FD grid with the SM calculus, displays small residual staircase diffractions''.

In this section, we will demonstrate our analysis and numerical tests for an interface at an angle of $22.5^\circ$. The analysis is then repeated for interfaces at angles of $0^\circ$, $11.75^\circ$, $22.5^\circ$, $33.75^\circ$ and $45^\circ$.

\subsection{Analysis methodology}
We will consider an interface that is positioned at an angle of $22.5^\circ$ away from the horizontal axis, as this angle maximizes the severity of staircase diffractions \citep{mittet2017}. A source will be excited at an offset of 200 meters to the interface. Seven horizontal particle velocity ($v_x^{sr}$) receivers are introduced. They are placed at 300 meters perpendicular to the interface and at 0, 100, \dots, 600 meters parallel to the interface. We will use the FD nodes closest to the described coordinates in practice, such that no interpolation is required. By avoiding interpolations, we may be certain that the recordings are accurate. We run simulations with $\Delta x=\Delta z=10$ m, 8 m, 6 m, 4 m, and 2 m. It is expected that for decreasing spatial steps, the accuracy should increase for all methods. The accuracy of the FD scheme, as described previously, is assumed to be excellent and free of source-generated and propagating (dispersive) errors. It is thus the assumption that any residual errors originate from the different interface representation methods.

The tested interfaces are acoustic-acoustic, elastic isotropic-elastic isotropic, and elastic aniso\-tropic-elastic anisotropic. The upper medium will contain the source and receivers, and is the slow medium. A stiffness matrix and constant density will be defined for this upper medium. The lower medium is then specified as 4.5 times the stiffness matrix and 1.5 times the density of the upper medium. The smallest velocities in all simulations are 1000-1200 m/s in the upper medium. We use a 17.5 Hz Ricker wavelet as the source-time function. The Ricker wavelet is not band-limited, but its amplitudes are negligible beyond about 45 Hz. Thus, with minimum velocities of 1000 m/s, we find a minimum wavelength of about 22 m. At the coarsest grid spacing of $\Delta x=10$ m, that gives 2.2 grid points per minimum wavelength. At the finest grid spacing of 2 m, that gives 11 grid points per minimum wavelength.

The analysis is then performed in three steps:
\begin{enumerate}
    \item We display snapshots for each of the simulations with the different methods for interface representation. The snapshots provide a visual overview of the source/receiver geometry and the expected wavefields. Additionally, they provide a qualitative check on the level of staircase diffractions.
    \item We display some traces of the reflection response as examples, excluding the direct wave. The direct wave was removed by running an additional simulation without an interface present and recording the direct wave at the receivers. We can then subtract this direct wave from the other measurements. Removing the direct wave (which is accurately modeled) means that we isolate the reflections (which may be inaccurate). Additionally, we will show reference solutions that are computed with the Cagniard-de Hoop method. For the acoustic and elastic isotropic case, we use the EX2DELEL code from \cite{berg1994analytical}. For the elastic anisotropic case, we have implemented the methodology described in \cite{hijdenthesis1987}. We compute the sample-by-sample differences between the FD solutions and reference solutions.
    \item To quantify the errors, we compute the $L^2$ normed error between the FD solution and each receiver's reference solution. We take the average of these errors over the seven receivers for each interface treatment method. We can then compare the relative accuracy of the different techniques. We plot the results for a linear scale of grid points per wavelength, thus using the reciprocal of the grid spacing. This generalizes the results for cases beyond those studied here.
\end{enumerate}

As a final technical detail, we note that the Cagniard-de Hoop reference solutions were computed for a horizontally stratified medium. To allow us to compare the simulated $v_x^{sr}$ field to the reference solution, we rotated (1) the source, (2) the receiver, and (3) the stiffness matrices in the reference solutions by $22.5^\circ$, such that the reference solution equals the set-up of the FD simulations.

\subsection{Acoustic-acoustic interface}
For the upper medium (containing the source and receivers), we use a density of $\rho=1000$ kg/m$^3$ and a stiffness matrix
\begin{equation}
    \begin{pmatrix} C_{11} & C_{13} & C_{15} \\ C_{13} & C_{33} & C_{35} \\ C_{15} & C_{35} & C_{55} \end{pmatrix}_\text{acoustic} = \begin{pmatrix} 1000\cdot 1200^2 & 1000\cdot 1200^2 & 0 \\ 1000\cdot 1200^2 & 1000\cdot 1200^2 & 0 \\ 0 & 0 & 0 \end{pmatrix}\ \text{Pa},
\end{equation}
resulting in a P-wave velocity of 1200 m/s. The lower medium has a stiffness matrix that is the above matrix multiplied with a factor 4.5, and a density of $1500$ kg/m$^3$. Effectively, that gives a P-wave velocity of about 2080 m/s in the lower medium. The source is excited as an explosive source in the center of the domain. As interface representations, we consider (1) sampling at the coarse resolution, (2) the anti-aliased step-function, (3) the 2-D lowpass filter, (4) the SM calculus. The orthorhombic averaging is identical to the SM calculus in this acoustic-acoustic setting, thus equals method (4). All simulations are run on a \cite{virieux1984sh} grid.

\begin{figure}
    \centering
    \includegraphics[width=0.85\textwidth]{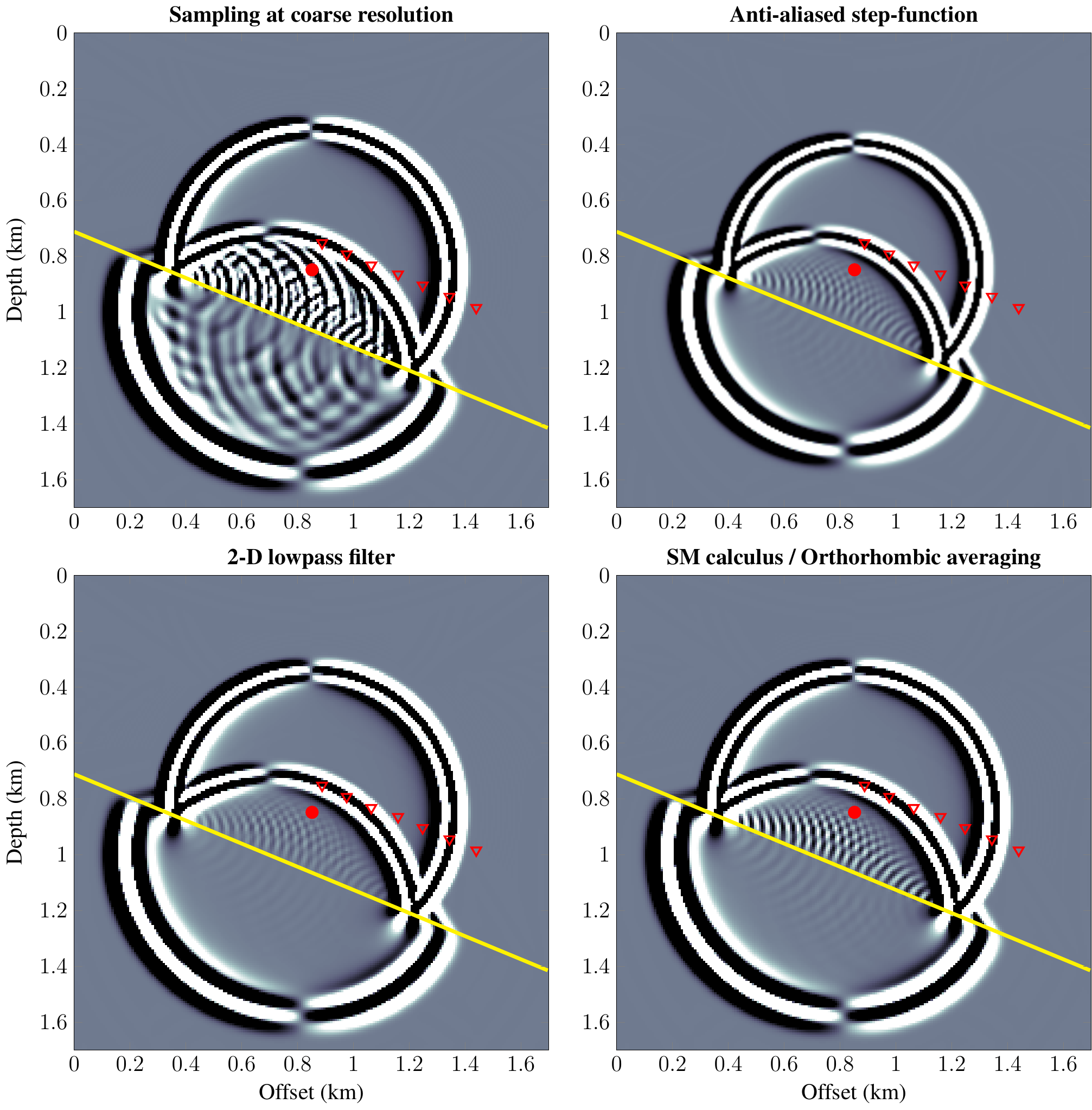}
    \caption{Horizontal particle velocity snapshots at $t=0.5$ s for the acoustic-acoustic interface with $\Delta x=\Delta z=8$ m. The filled red circle represents the source position. The red triangles represent the receiver positions. The snapshots have been clipped to an identical level of 1\% of the maximum amplitude.}
    \label{fig:snapshots_acoustic}
\end{figure}

Snapshots of the wavefield due to the acoustic-acoustic interface with $\Delta x=\Delta z=8$ m are shown in Figure \ref{fig:snapshots_acoustic}. A direct P-wave is visible. Additionally, a reflected wave, a transmitted wave, and a head wave component can be seen. Numerical artifacts in the form of curved lines around the interface can also be seen. These staircase diffractions are most visible with sampling at the coarse resolution. Use of the SM calculus (and thus also the orthorhombic averaging) already significantly reduces the staircase diffractions. The artifacts are, however, smallest with the anti-aliased step-function and the 2-D lowpass filter.

\begin{figure}
    \centering
    \includegraphics[width=0.8\textwidth]{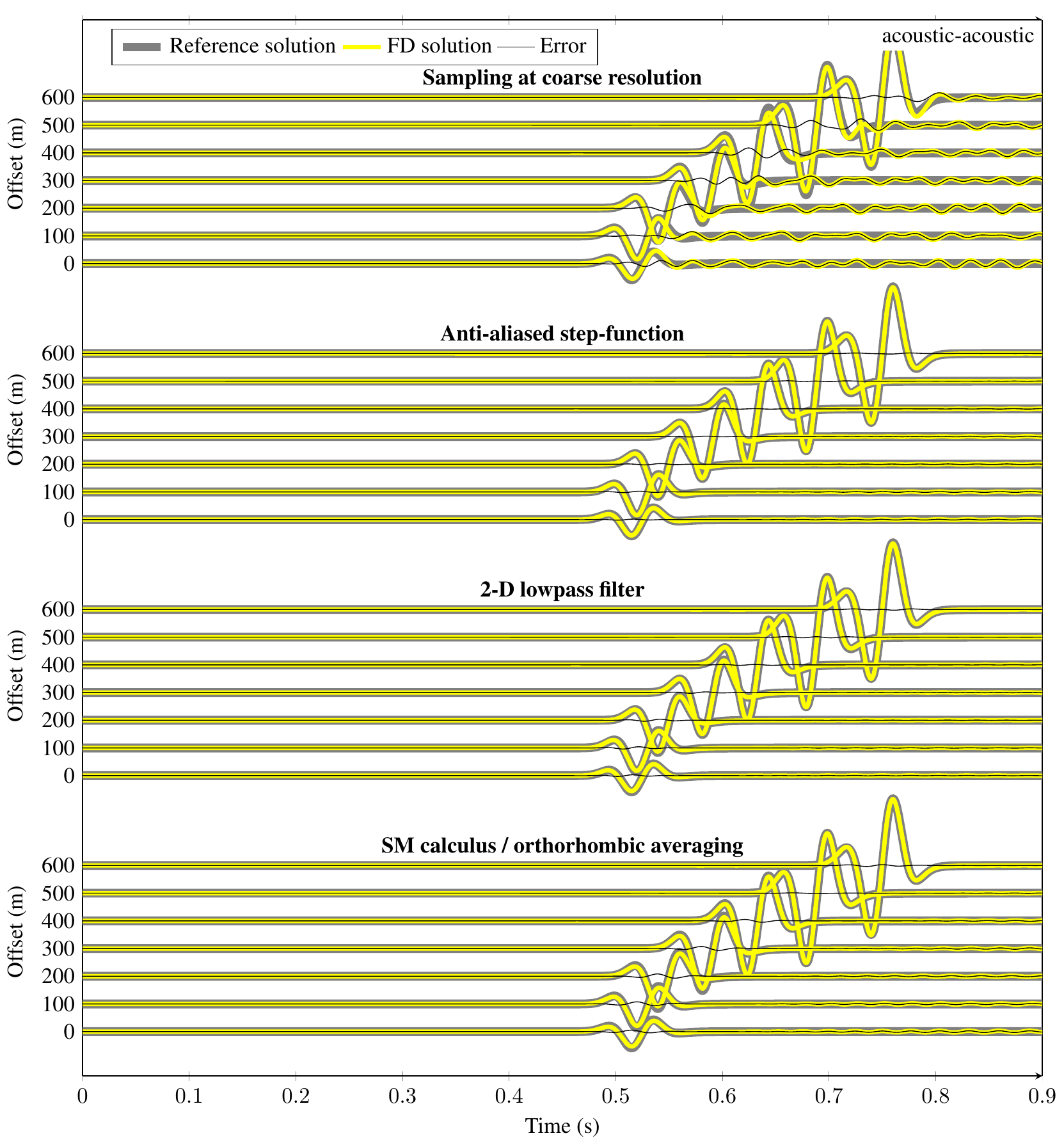}
    \caption{An example of the computed $v_x^{sr}$ traces for the acoustic-acoustic interface, with $\Delta x=\Delta z=8$ m. The traces have been offset vertically to represent the different receiver offsets.}
    \label{fig:traces_acoustic}
\end{figure}

In Figure \ref{fig:traces_acoustic}, some example traces are provided with $\Delta x=\Delta z=8$ m. The recorded traces are compared to reference solutions. The first method, sampling at the coarse resolution, shows clear rippling artifacts that follow the main reflections. The ripples are caused by the staircase diffractions. The rippling artifacts are much smaller in the other three remaining methods. The 2-D lowpass filter method creates the smallest amount of residual ringing. Notably, the SM calculus (and thus the orthorhombic averaging) result in amplitude mismatches that are larger for small offsets than for far offsets. No such amplitude-versus-offset error can be seen for the two anti-aliasing strategies.

\begin{figure*}
    \centering
    \includegraphics[width=1\textwidth]{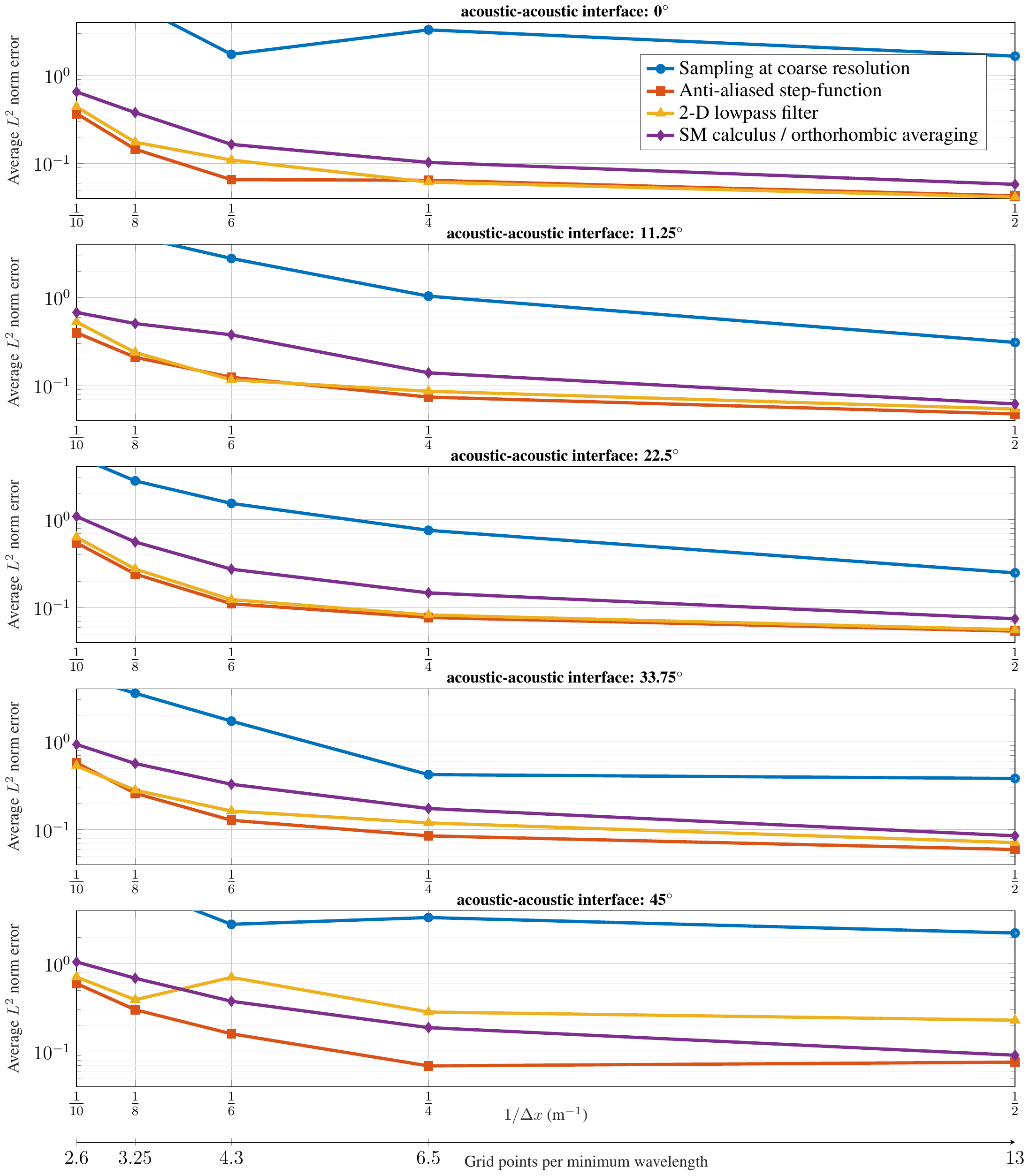}
    \caption{The results for interfaces dipping at various angles, for an acoustic-acoustic interface.}
    \label{fig:acoustic_gpw}
\end{figure*}

Figure \ref{fig:acoustic_gpw} displays the average $L^2$ error for all receivers, all used step sizes, and interfaces at five different angles (not just the $22.5^\circ$ case that was considered thus far). As expected, decreasing the step size (towards the right) decreases the errors. Additionally, we see that the first method (sampling at the coarse resolution) results in the largest errors. In comparison, the two anti-aliasing methods generally yield the smallest errors. We may also note that the $L^2$ error for sampling at the coarse resolution does not decrease monotonically for the cases of interfaces at $0^\circ$ and $45^\circ$, for unknown reasons. The SM calculus (and thus the orthorhombic averaging) yields errors that are greater but not far from those of the anti-aliasing methods. Note that the largest gains in accuracy (nearly an order of accuracy in error reduction) are made by moving from 2.6 to 4.3 grid points per minimum wavelength. In this range, the anti-aliasing methods reach equivalent accuracy to the SM calculus with 1.5 times larger spatial steps. An exception is found for the $45^\circ$ case, where the 2-D lowpass ﬁlter causes worse results than the anti-aliased step-function and SM calculus. This larger error may caused by the windowed Hanning filter, which may not be sufficient for the given data. Regardless, the anti-aliasing methods generally yield more accurate results than the equivalent media methods.

\subsection{Elastic isotropic-elastic isotropic interface}
For the upper medium (containing the source and the receivers), we use a density of $\rho=1000$ kg/m$^3$ and a stiffness matrix
\begin{equation}
    \begin{split}&\begin{pmatrix} C_{11} & C_{13} & C_{15} \\ C_{13} & C_{33} & C_{35} \\ C_{15} & C_{35} & C_{55} \end{pmatrix}_\text{isotropic} =\\
    &\begin{pmatrix} 1000\cdot 2000^2 & 1000\cdot2000^2-2\cdot 1000\cdot1200^2 & 0 \\ 1000\cdot2000^2-2\cdot 1000\cdot1200^2 & 1000\cdot 2000^2 & 0 \\ 0 & 0 & 1000\cdot 1200^2 \end{pmatrix}\ \text{Pa},
    \end{split}
\end{equation}
resulting in a P-wave velocity of 2000 m/s and an S-wave velocity of 1200 m/s. The lower medium has a stiffness matrix that is the above matrix multiplied with a factor 4.5, and a density of 1500 kg/m$^3$, such that lower medium has a P-wave velocity of nearly 3500 m/s and an S-wave velocity of just over 2000 m/s. The source is excited as an explosive source in the center of the domain. As interface representations, we consider (1) sampling at the coarse resolution, (2) the anti-aliased step-function, (3) the 2-D lowpass filter, (4) the SM calculus, (5) the orthorhombic averaging. Note that for the SM calculus, we must carry out an anisotropic simulation on a \cite{lebedev1964difference} grid. The other methods can be modeled using an orthorhombic modeling scheme on a \cite{virieux1986} grid. The simulations for the SM calculus thus require twice as much computational effort.

\begin{figure}
    \centering
    \includegraphics[width=0.85\textwidth]{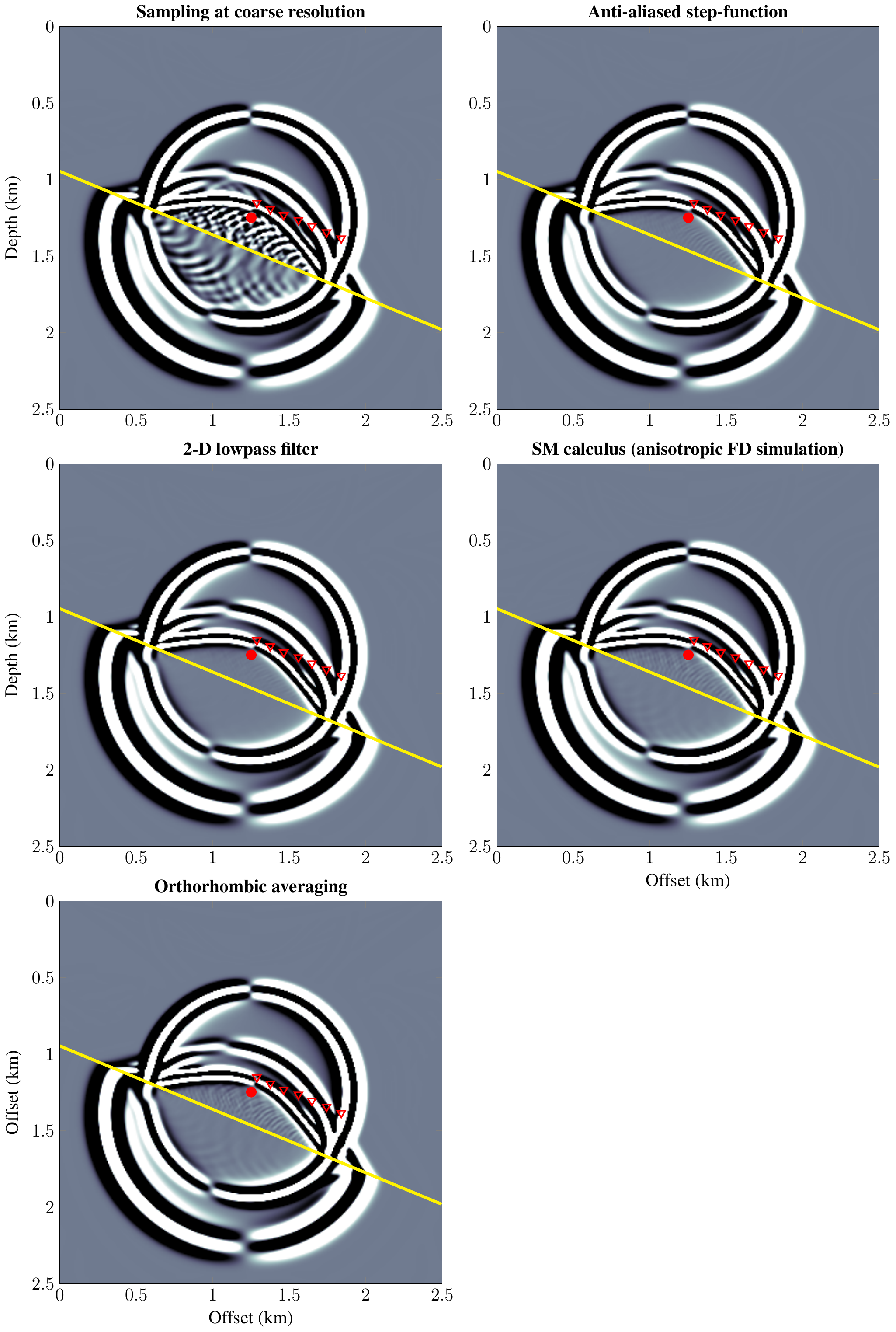}
    \caption{Horizontal particle velocity snapshots at $t=0.5$ s for the elastic isotropic-elastic isotropic interface with $\Delta x=\Delta z=8$ m. The filled red circle represents the source position. The red triangles represent the receiver positions. The snapshots have been clipped to an identical level of 1\% of the maximum amplitude.}
    \label{fig:snapshots_isotropic}
\end{figure}

Snapshots of the wavefield due to the elastic isotropic-elastic isotropic interface with $\Delta x=\Delta z=8$ m are presented in Figure \ref{fig:snapshots_isotropic}. The direct P-wave is again visible. Compared to the previous case, we now observe an additional reflected S-wave. Additionally, four head wave reflections are generated. The staircase diffractions are most obvious with sampling at the coarse resolution. The SM calculus and orthorhombic averaging create small residual amounts of staircase diffractions. The staircase diffractions are still smallest for the two anti-aliasing methods. 

\begin{figure}
    \centering
    \includegraphics[width=0.8\textwidth]{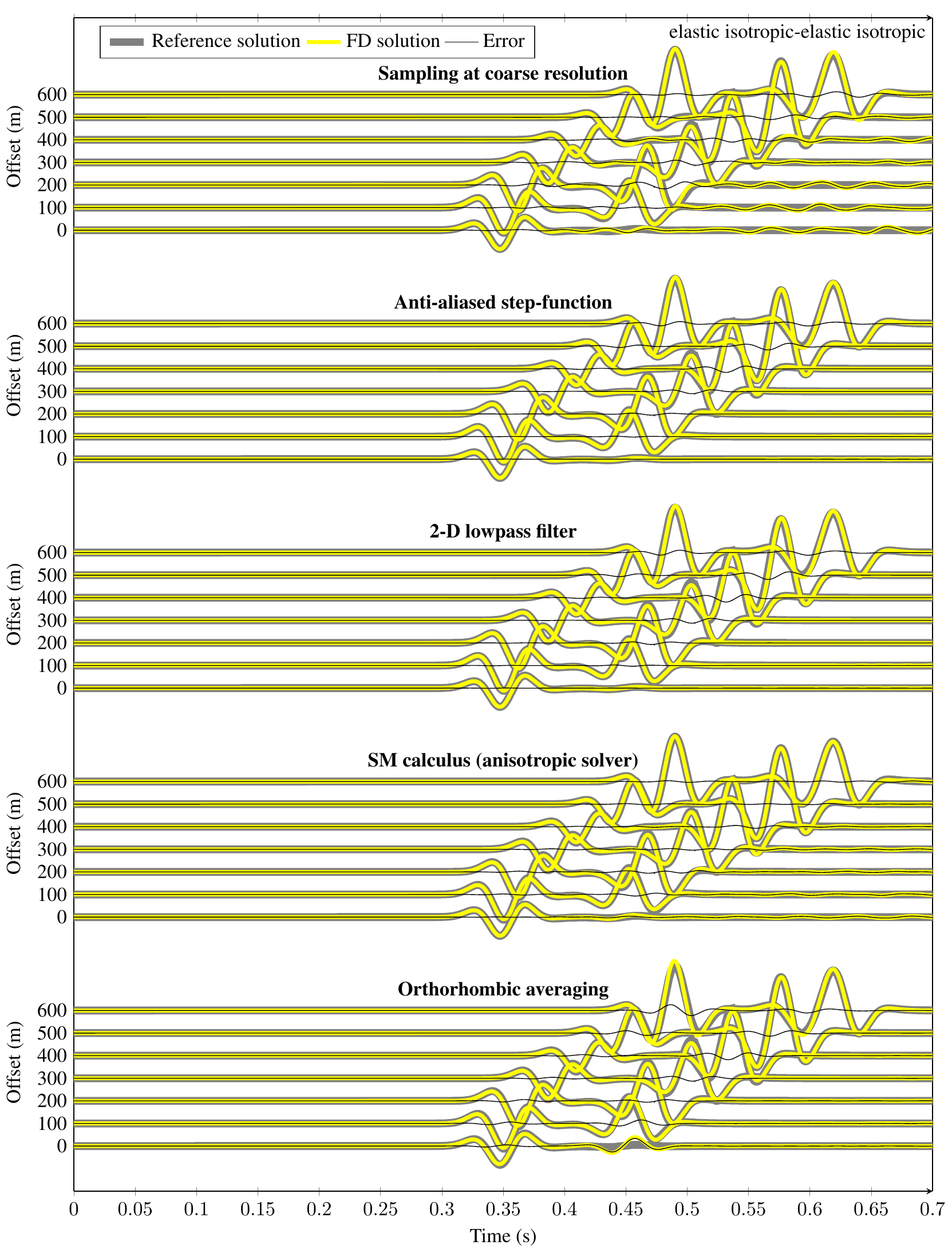}
    \caption{An example of the computed $v_x^{sr}$ traces for the elastic isotropic-elastic isotropic interface, with $\Delta x=\Delta z=8$ m. The traces have been offset vertically to represent the different receiver offsets.}
    \label{fig:traces_isotropic}
\end{figure}

In Figure \ref{fig:traces_isotropic}, some example traces are provided with $\Delta x=\Delta z=8$ m. The recorded traces are compared to reference solutions. The first method (sampling at the coarse resolution) shows clear rippling artifacts that follow the main reflections and refractions. The ripples are caused by the staircase diffractions. The rippling artifacts are much smaller in the other four remaining methods; the two anti-aliasing methods and orthorhombic averaging yield the smallest levels of staircase diffractions. The orthorhombic averaging, however, contains numerous phase and amplitude errors. For example, at the zero offset, the reflected S-wave has a too large amplitude. The two anti-aliasing methods create their own notable set of artifacts. It is clear that for the zero offset recordings, the amplitudes and phases are correct. For larger offsets, however, the amplitudes and phases become erroneous. The field computed with the SM calculus (thus computed with an anisotropic solver) is slightly incorrect for the small offsets but very accurate for the large offsets. The smallest error levels are thus computed with the SM calculus, despite the higher levels of staircase diffractions.

\begin{figure*}
    \centering
    \includegraphics[width=1\textwidth]{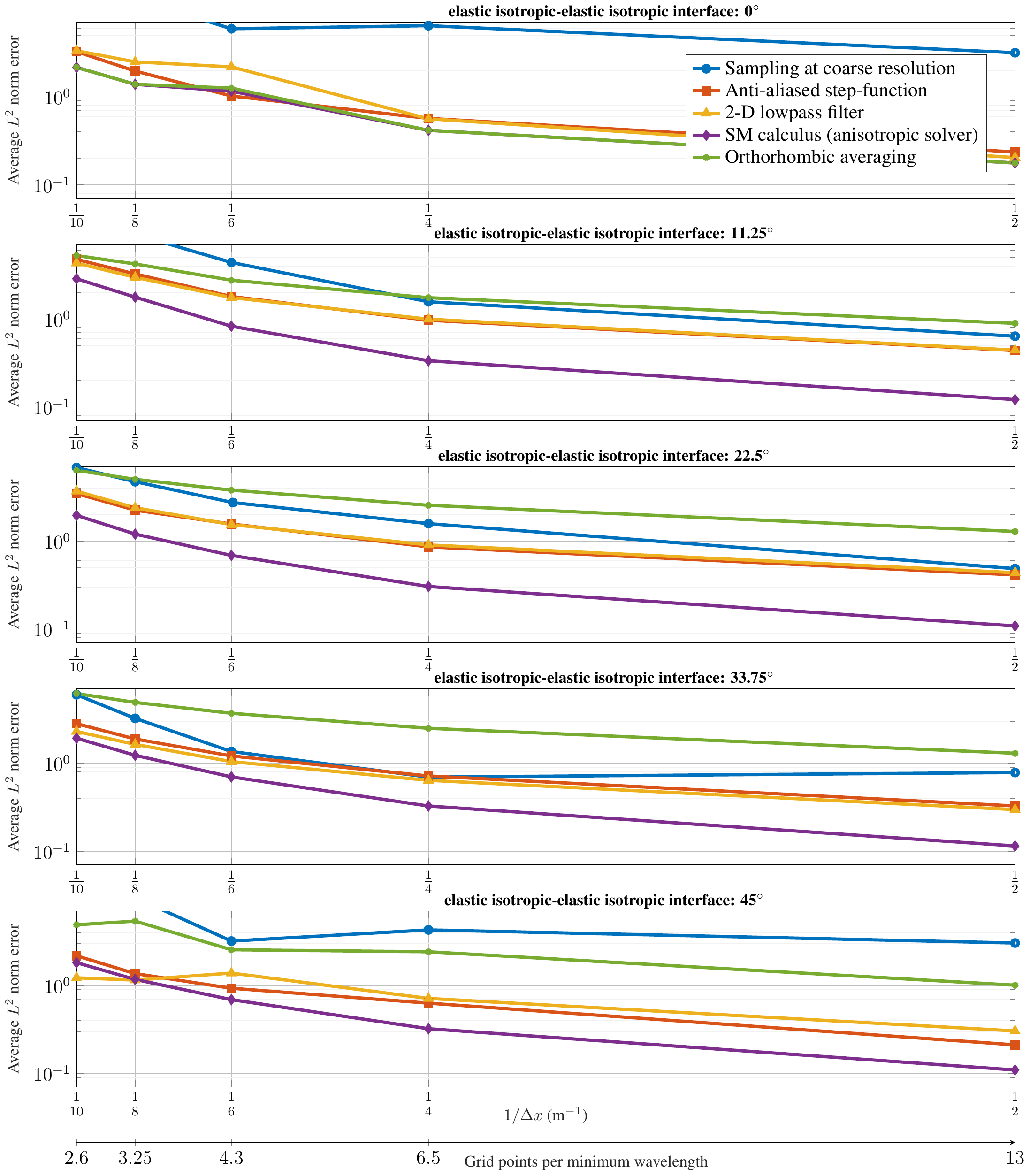}
    \caption{The results for interfaces dipping at various angles, for an elastic isotropic-elastic isotropic interface.}
    \label{fig:isotropic_gpw}
\end{figure*}

Figure \ref{fig:isotropic_gpw} displays the average $L^2$ error for all used receivers, all used step sizes, and all five angles (not just the $22.5^\circ$ case that was considered thus far). Decreasing the step size generally reduces the errors. We note that for interface angles dipping at $0^\circ$ and $45^\circ$, the decrease is not monotonous. At $0^\circ$ the entire interface lies at a similar depth level with respect to the grid, and then varying the grid cell size may place the interface in a more or less error-inducing location. Conversely, for the $45^\circ$ case, it is not clear why the results are not monotonously decreasing. We can see that for a flat interface at $0^\circ$, the SM calculus and orthombic averaging give the best results -- they are in principle identical for a flat interface, the small differences in the computed traces stem from the difference in establishing relative volumes of the two media, as discussed in the Theory section. For interfaces at other angles, however, the orthorhombic averaging is not as accurate. On the other hand, the SM calculus (using an anisotropic solver) produces the smallest errors for interfaces at essentially all angles. The two anti-aliasing methods have intermediate levels of errors. The SM calculus achieves equivalent accuracy with up to 1.25 times larger steps compared to the anti-aliasing methods.


\subsection{Anisotropic elastic-anisotropic elastic interface}
For the upper medium (containing the source and the receivers), we use a density of $\rho=2000$ kg/m$^3$ and a stiffness matrix
\begin{equation}
    \begin{pmatrix} C_{11} & C_{13} & C_{15} \\ C_{13} & C_{33} & C_{35} \\ C_{15} & C_{35} & C_{55} \end{pmatrix}_\text{anisotropic} = \begin{pmatrix} 15.6 & 7.7 & -3.4 \\ 7.7 & 14.3 & 0.1 \\ -3.4 & 0.1 & 4.3 \end{pmatrix} \ \text{GPa}.
\end{equation}
We rotate the medium by $22.5^\circ$ using, e.g., a Bond transform. For the simulations at other angles, we also rotated the stiffness matrix by the respective angle of the dipping interfaces, to make sure we model the exact same reflecting and refracting events in each simulation. The stiffness matrix and density correspond to a minimum qP-wave group velocity of just over 2300 m/s, and a minimum qS-wave group velocity of 1000 m/s. The lower medium has a stiffness matrix that is the above matrix multiplied with a factor 4.5, and a density of 3000 kg/m$^3$. The lower medium has a minimum qP-wave group velocity of nearly 6000 m/s and a minimum qS-wave group velocity of just over 2100 m/s. The source is excited as a horizontally directed force-source, in the center of the domain. As interface representations, we consider (1) sampling at the coarse resolution, (2) the anti-aliased step-function, (3) the 2-D lowpass filter, (4) the SM calculus. The orthorhombic averaging does not apply to anisotropic media, thus is not considered. All the simulations are carried out on a \cite{lebedev1964difference} grid.

\begin{figure}
    \centering
    \includegraphics[width=0.85\textwidth]{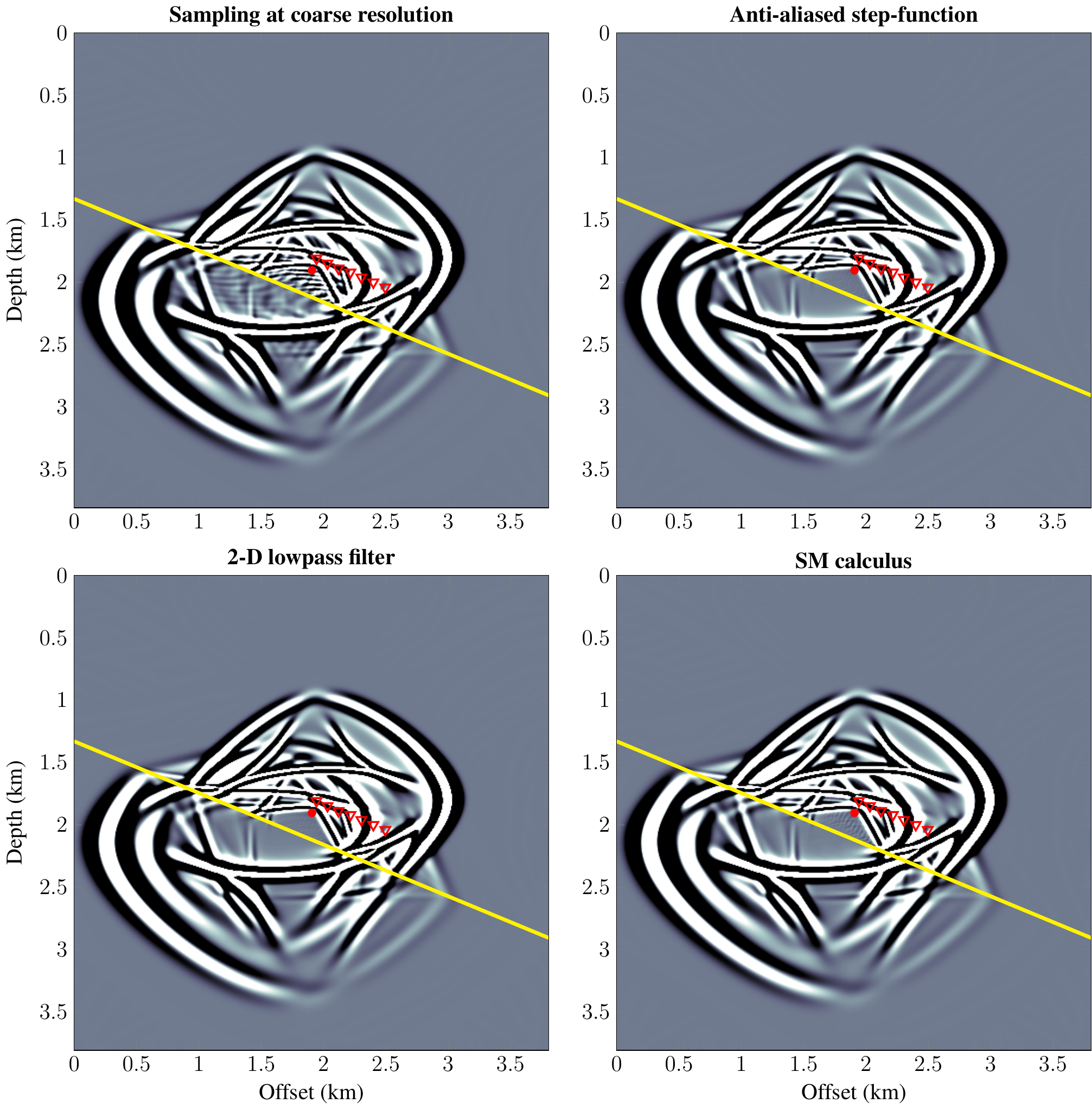}
    \caption{Horizontal particle velocity snapshots at $t=0.44$ s for the elastic anisotropic-elastic anisotropic interface with $\Delta x=\Delta z=8$ m. The filled red circle represents the source position. The red triangles represent the receiver positions. The snapshots have been clipped to an identical level of 1\% of the maximum amplitude.}
    \label{fig:snapshots_anisotropic}
\end{figure}

Snapshots of the wavefield due to the elastic anisotropic-elastic anisotropic interface with $\Delta x=\Delta z=8$ m are provided in Figure \ref{fig:snapshots_anisotropic}. We observe that the wavefield is relatively complicated in this case: qP and qS wavefronts are generated, together with a multitude of reflections and head waves. The staircase diffractions are most apparent with the sampling at the coarse resolution. The SM calculus leads to residual staircase reflections upon closer inspection. Conversely, the two anti-aliasing methods lead to no discernible staircase diffractions.

\begin{figure}
    \centering
    \includegraphics[width=0.8\textwidth]{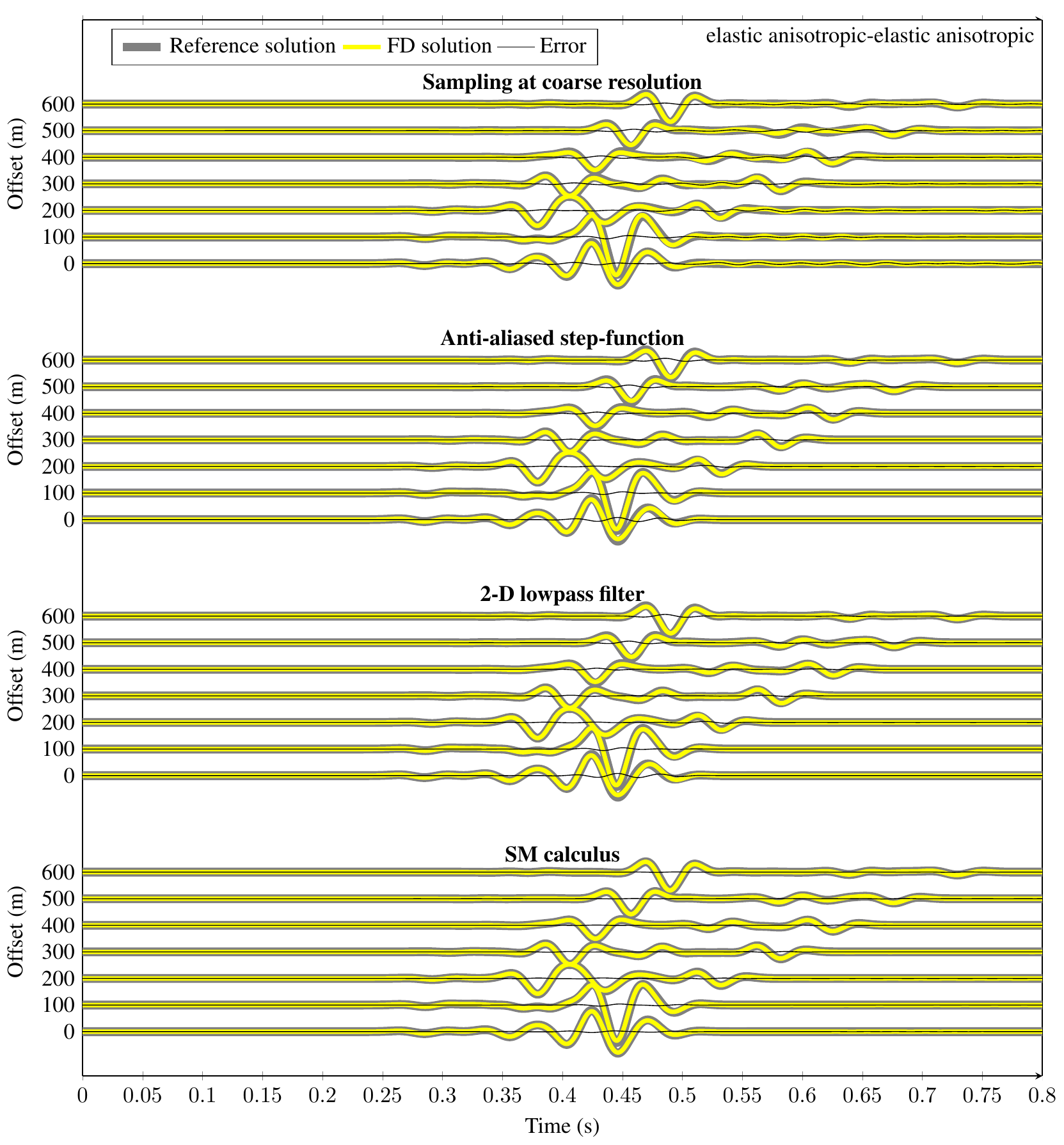}
    \caption{An example of the computed $v_x^{sr}$ traces for the elastic anisotropic-elastic anisotropic interface, with $\Delta x=\Delta z=8$ m. The traces have been offset vertically to represent the different receiver offsets.}
    \label{fig:traces_anisotropic}
\end{figure}

In Figure \ref{fig:traces_anisotropic}, some example traces are provided with $\Delta x=\Delta z=8$ m. The first method (sampling at the coarse resolution) shows clear rippling artifacts that follow the main events, caused by staircase diffractions. Such rippling artifacts are not visible with the other methods considered. We further note that the two anti-aliasing methods give results with slightly erroneous phases and amplitudes. This holds for receivers at both small and large offsets. The SM calculus may have slightly erroneous amplitudes at small offsets, but produces excellent phase and amplitude fits at long offsets.

\begin{figure*}
    \centering
    \includegraphics[width=1\textwidth]{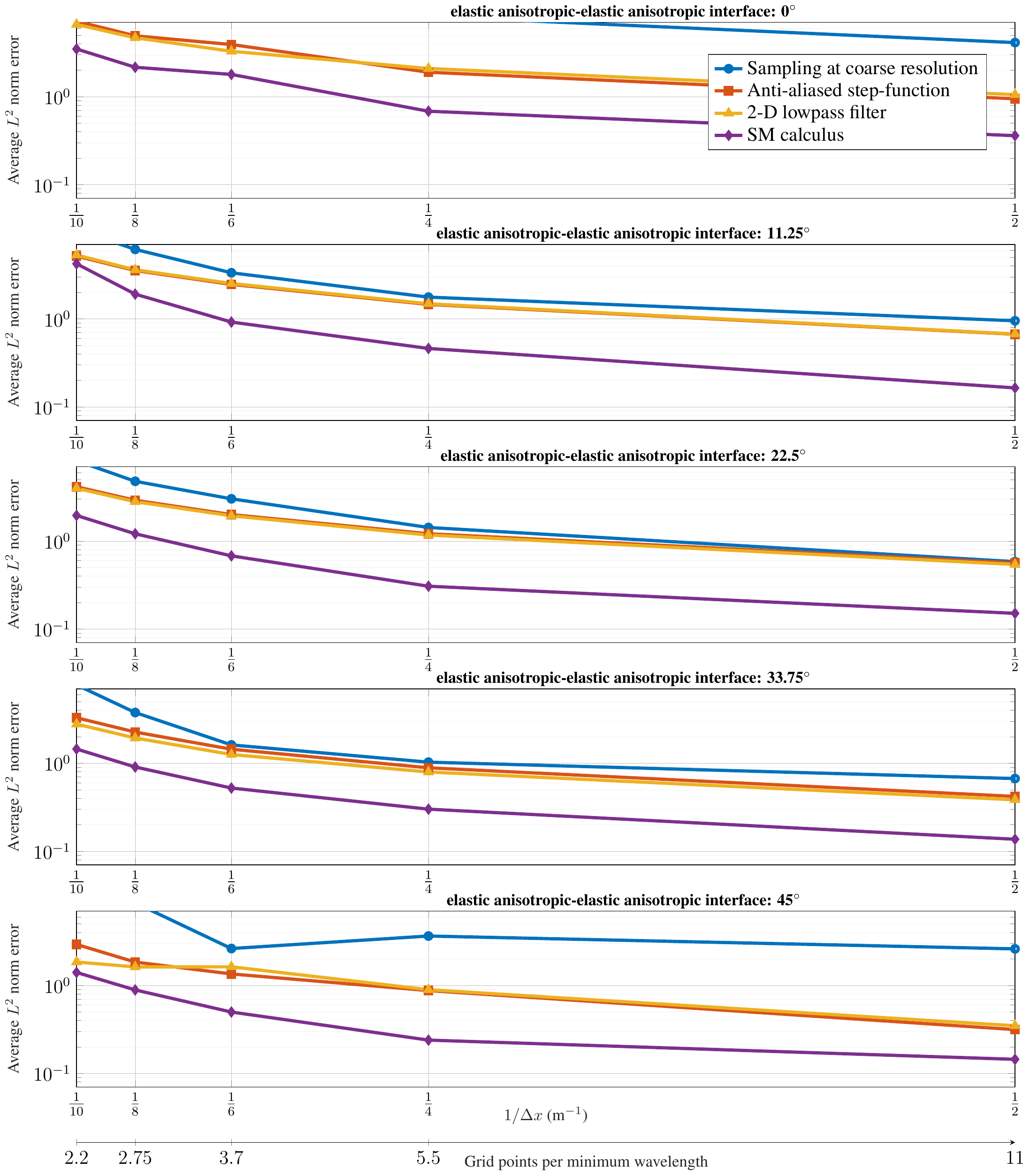}
    \caption{The results for interfaces dipping at various angles, for an elastic anisotropic-elastic anisotropic interface.}
    \label{fig:anisotropic_gpw}
\end{figure*}

Figure \ref{fig:anisotropic_gpw} displays the average $L^2$ error for all used receivers, all used step sizes, and all five interface angles (not just the $22.5^\circ$ case that was considered thus far). Decreasing the step size generally reduces the errors. The only exception is the sampling at a coarse resolution and 2-D lowpass filter, which at $45^\circ$ do not have monotonously decreasing errors, for reasons unknown. We can see that the sampling at coarse resolution produces the largest errors. On the other hand, the SM calculus creates the highest level of accuracy. The two anti-aliasing methods produce intermediate levels of errors. For smaller grid spacings, the anti-aliasing methods are seemingly no better than sampling at a coarse resolution. The SM calculus thus overall clearly achieves the best performance.


\section{Discussion}
\subsection{The problem set-up}
In the examples, we have used a large stencil size of $L=20$, to make sure the results were virtually free of spatial dispersion. However, one may reasonably object that this is a stencil size much larger than what is ever used in practice, and that perhaps also the results would be different if we use a smaller stencil size. For this reason, we have repeated the analysis with a stencil size of half-order $L=3$. One result for the elastic isotropic-elastic isotropic case at $22.5^\circ$ is presented in Appendix \ref{app:smallerFDstencilresults}. It can be seen that the general trend is identical to the previously obtained results that were presented in Figure \ref{fig:isotropic_gpw}, other than that the errors of \textit{all} methods are slightly increased. The latter increase in errors is likely due to the increased levels of numerical dispersion. Hence, it is not expected that the stencil size has played a significant role in our analysis.

In the examples, we have considered a very particular and perhaps contrived test-case for reflections off straight (i.e., not-curved) interfaces separating two homogeneous half-spaces. We limited ourselves to this set-up because we could test the accuracy of the solution against analytical solutions that exist for layered media. Our results indicate that a high-resolution (i.e., $\Delta x=\Delta z=2$ m) simulation where we sample at the coarse resolution would not have been an appropriate reference simulation, as the errors in such a naive method would occasionally have been larger than the errors obtained by other methods at much coarser resolution (i.e., $\Delta x=\Delta z=10$ m). It seems that simply using fine FD simulations would be a prohibitively expensive method to obtain sufficiently accurate reference results, even in this simple set-up. Hence, if we want to consider tests as done in this paper for more complicated interface set-ups, such as reflections off curved interfaces, we will struggle to find appropriate reference solutions to test the accuracy of our schemes. Other modeling methods better suited to complicated model geometries, such as spectral element modeling methods, may perhaps provide a way forward. This was, for example, done in \cite{kristekorthorhombic2016}. To conclude, our conclusions apply first and foremost to the simple set-up considered here, and it is neither clear nor easily tested how well the results apply to other heterogeneous domains. 

\subsection{Extension to multiple interfaces}
In the examples discussed, we have considered only the case of a single interface. This limitation is in place such that we can carefully compare results against analytical solutions. It is generally trivial to extend the methods presented here to the case of multiple interfaces as long as these interfaces do not cross. Conversely, it is not trivial for some of the methods to deal with interfaces that converge to a point (for example, a thin wedge model). Only the 2-D lowpass filtering method and the orthorhombic averaging method will be able to deal with the presence of two interfaces within a single Voronoi FD cell, as these methods only depend on supersampling the rock property model. Conversely, the anti-aliased step-function and SM calculus can only evaluate the rock properties along a single interface, as they explicitly depend on the dipping angle of this single interface. Hence, the 2-D lowpass filtering method and the orthorhombic method have significantly more flexibility when it comes to constructing models with arbitrary interfaces or levels of sub-cell model variations. This makes the latter two methods easy to put into practice with the sidenote, however, that these methods may not always provide the best levels of accuracy, as shown in the results section.


\subsection{The $L^2$ cost function}
Quantifying the errors with the $L^2$ error, averaged over seven receivers, biases the results. Both the averaging procedure and $L^2$ norm are, for example, somewhat sensitive to outliers. A single poor FD result at one receiver can thus skew the results. Furthermore, the $L^2$ norm is more susceptible to phase errors than to amplitude errors: ``\emph{The standard RMS \emph{[i.e., $L^2$]} misfit matches the single-valued envelope misfit only in the case of a pure amplitude modification of the signal. In all other cases RMS considerably overestimates the misfits and does not characterize them.}'' \citep{kristekova2006misfit}. A different cost function could thus lead to a different conclusion. For example, one could separately weigh the phase and amplitude errors. \cite{kristekova2006misfit} and \cite{mittet2017} provide examples of such more involved error functions. However, to compress the large amount of data into a few numbers, we opted for the $L^2$ normed error. Furthermore, the $L^2$ norm is a standard measure of inaccuracy in many geophysical inversions. Therefore, creating error-free synthetics in an $L^2$ sense is an important feature in and by itself.

Although not obvious from Figure \ref{fig:snapshots_isotropic} and Figure \ref{fig:traces_isotropic}, the orthorhombic averaging shows up as an inferior method to sampling at a coarse resolution in the isotropic case as shown in Figure \ref{fig:isotropic_gpw}. Despite the fact that the orthorhombic averaging removes many artifacts that occur in the coarse resolution model, it results in amplitude and phase errors for all offsets. Some of these errors are relatively large (as measured with the $L^2$ error). Conversely, the sampling at a coarse resolution has many errors, but those errors have a smaller magnitude (again, as measured with the $L^2$ error). In some sense, the inferior performance of the orthorhombic averaging compared to sampling at a coarse resolution is thus misleading, and may be ascribed to the use of the $L^2$ normed error.

\subsection{Accuracy considerations}
For the two best-performing techniques, the anti-aliasing methods and the SM calculus, we see two interesting and opposing trends:
\begin{enumerate}
\item The anti-aliasing methods are accurate at small offsets, but erroneous for long offsets. Hence, they are accurate in 1-D when only reflections occur. But they are less accurate in 2-D when reflections but also mode conversions and head waves occur. The origin of these errors might be found in the distributed nature of the anti-aliasing methods. That is, the interface is spread over multiple FD cells. The Gibbs phenomenon around the interface (i.e., the overshooting) leads to both smaller and larger phase velocities compared to the fine velocity model. A wave that hits an interface head-on thus interacts with only a few of the anti-aliased FD cells. Conversely, a wave that hits the interface at a more oblique angle interacts with many anti-aliased FD cells. An oblique wave may then take on an erroneous group velocity when propagating along the interface. It is feasible to assume that this leads to phase and amplitude errors in the simulation at longer offsets.
\item 
The SM calculus is (slightly) incorrect for small offsets, but very accurate for long offsets. The reason for this may also be related to the above explanation. That is, a wave that hits the equivalent medium interface head-on may lead to errors. After all, these are the type of errors that the anti-aliasing methods fix by distributing the interface over multiple FD cells. On the other hand, oblique waves will propagate parallel to the interface with the correct group velocity. At a more oblique angle, furthermore, the wave interacts with several FD cells before reflecting or refracting upwards. Oblique waves thus encounter a distributed interface naturally. The errors then remain small along the interface.
\end{enumerate} 
The explanations given above are only hypotheses. The ad-hoc nature of both anti-aliasing methods and SM calculus does not allow for proofs regarding their accuracy in FD simulations. However, the trends were found to hold for many tests that we performed, with both inclined and flat interfaces.

The fact that anti-aliasing outperforms the SM calculus in the acoustic case, but not in the elastic case, is interesting. We may speculate as to its origin in three ways:
\begin{enumerate}
    \item The SM calculus in the acoustic case does not become anisotropic (unlike the elastic case), thus cannot produce the correct angle-dependent group velocities;
    \item The physics in the acoustic case are sufficiently simple (e.g., there is no angle-dependent mode conversion between P and S waves) that the anti-aliasing methods (as discussed in the previous section) do not produce serious errors;
    \item Alternatively, it may be possible that for the elastic case, we are simply anti-aliasing the erroneous FD quantities. The fact that the density and compliance matrix are filtered is based on ad-hoc reasoning in \cite{mittet2017} that works very well in acoustic data, but it cannot be proven from first principles. Perhaps an alternative set of quantities may lead to more accurate results also in the elastic cases.
\end{enumerate}
Again, with a lack of a theoretical framework explaining the origin of the FD errors in the presence of interfaces, one can only speculate which of these explanations is correct.

\subsection{Computational cost}
Finally, a word on the computational cost of the interface representation methods. The equivalent media methods fall in the class of embarrassingly parallel tasks \citep{hobro2010rapid}. That is, we can compute the equivalent medium for one FD cell independently from all other FD cells. Thus, computing the SM calculus for intersected FD cells is a task that is easily carried out in parallel. The nature of the 2-D lowpass filter is quite different, because any FD cell is affected by neighboring cells. However, a windowed lowpass filter operation can still be parallelized. For example, the domain can be split into two and processed simultaneously, allowing for some overlap for the filter windows in both parts. This splitting procedure may be repeated a few times. Regardless, the 2-D lowpass filter had a low cost in our tests (both in runtime and memory overhead). Following the simplifications in eqs. \eqref{eq:acousticcompliancereinverted} and \eqref{eq:isotropiccompliancereinverted}, we only need to lowpass filter one or two grids, and then combine those grids to have a final grid ready for computations. In the tests done here, such grids could be computed within two seconds to produce the entire filtered velocity model. Hence the anti-aliasing methods are cheap for acoustic and isotropic media. Their cost was negligible compared to the FD simulations, which could take up to six hours.

The SM calculus leads to the best results in 2-D elastic media. But the requirement of an anisotropic solver, just to model an interface between two isotropic media, also doubles the computational cost. Ideally, one thus uses an anisotropic solver around the interface only, and an isotropic solver elsewhere. However, such hybrid methods are not trivial to make error-free. For example, \cite{lisitsa2012numerical} show that mixing the Virieux and Lebedev grids requires 20-50 grid points per minimum wavelength to obtain errors of less than 0.3\%. The simulations carried out here are in the range of 2-11 grid points per minimum wavelength. So an alternative question arises: are the other interface representations more efficient, taking into account that they do not require an anisotropic FD solver? Figure \ref{fig:isotropic_gpw} reveals that the accuracy of the SM calculus may also be attained with the anti-aliasing methods, using about 20\% smaller grid spacings. So, which of the methods is more computationally attractive? We can make some estimates to answer this question, based on the scaling of the FD method. To use the anti-aliasing methods, we require $1/0.80\approx 1.25$ as many cells in each Cartesian direction and must reduce $\Delta t$ to satisfy the CFL criterion for smaller spacings. Hence, the increase in computational cost to run a simulation from $\Delta x=10$ to $\Delta x=8$ m is $(1/0.8)^3=1.95$, thus about twice as high. Hence, equivalent accuracy comes with an identical cost for both the SM calculus and the anti-aliasing methods. The SM calculus is, however, conceptually more straightforward. Furthermore, the SM calculus does not suffer from overshoot artifacts that can make simulations unstable. The SM calculus with an anisotropic solver is, thus, best suited for minimizing FD errors in elastic media, even in the isotropic case. 

If the errors of 3-D anti-aliasing and SM calculus are comparable to the errors in 2-D, it might be the case that the anti-aliasing methods can provide a more efficient solution in 3-D. This would be because the computational cost of the \cite{lebedev1964difference} grid FD solvers is four times larger than that of isotropic FD solvers in 3-D. If anti-aliasing results in an equivalent accuracy if the simulation is made finer by a factor 0.8, the cost increases only by a factor $(1/0.8)^4=2.4$. However, we have not yet carried out a similar analysis for 3-D media, thus do not know how the errors scale in that setting. This will be a subject of future research.

\section{Conclusion}
The modeling of inclined interfaces in coarse grid FD simulations is discussed. If the velocity model is naively sampled onto the coarse FD grid, significant errors appear. These errors take on the form of amplitude and phase errors, as well as staircase diffractions. We tested ways to sample the high-resolution velocity model onto a coarse FD grid to avoid such errors. These methods fall into two categories. The first category of methods uses anti-aliasing principles to smoothen the sharp interface using multiple grid points. The second category of methods uses equivalent media theory to replace intersected FD cells with an appropriately averaged medium. All techniques succeeded in the suppression of staircase diffractions. However, differences appeared in the reduction of phase and amplitude errors. In the acoustic case, the anti-aliasing approach resulted in the smallest error levels. In the elastic case, the SM calculus led to the smallest errors. A downside of the SM calculus in isotropic media is that it requires an anisotropic solver, which doubles the computational cost. However, the other methods also become twice as expensive to reach an equivalent level of accuracy. Hence, anisotropic modeling with SM calculus is an efficient method to minimize errors in elastic simulations with inclined interfaces in 2-D.

\section*{Data availability}
The MATLAB code underlying this article will be shared on reasonable request to the first author.

\section*{acknowledgments}
This work was supported by SNF grant 2-77220-15 and supported by The Swedish Research Council grant 2019-04878. We thank Dirk-Jan van Manen, Jozef Kristek, and James Hobro for interesting discussions. We gratefully acknowledge the work of Ludovic Métivier, Peter Moczo, and an anonymous reviewer, for their helpful and constructive comments to improve this manuscript.

\bibliographystyle{apa}
\bibliography{bibliography}

\appendix

\section{Anti-aliased step-function}\label{app:antialiasedstepfunction}
The Fourier transform of the Heaviside function in eq. \eqref{eq:heavisideinspace} is not well-defined (since the Heaviside function is not integrable), but may be taken in the language of distribution theory. For this purpose, we first define the Fourier transform and its inverse as a parameterized operator by $K$,
\begin{alignat}{5}\label{eq:forwardfourier}
    \widehat{p}(k) &=& \mathcal{F}[p](k) & = \int_{-\infty}^\infty p(z)e^{-ikz}\di z,& \\
    p_K(z)&=&\mathcal{F}^{-1}_{K}[\widehat{p}](z) & = \frac{1}{2\pi}\int_{-K}^K \widehat{p}(k)e^{ikz}\di k,&\label{eq:inversefourier}
\end{alignat}
where we may recognize the traditional set of Fourier transforms if we were to use $\mathcal{F}^{-1}_{\infty}$ for the inverse transform. The Fourier transform of the Heaviside function is known from distribution theory to be
\begin{equation}
    \widehat{H}(k) = \mathcal{F}[H](k) = \frac{1}{ik} + \pi \delta(k),
\end{equation}
where $1/(ik)$ is interpreted in terms of the Cauchy principal value, and $\delta(k)$ is the Dirac delta function. When we want to compute this Heaviside function only up to some limited wavenumber $K$, we can take a band-limited inverse Fourier transform,
\begin{equation}
    H_{K}(z) = \mathcal{F}^{-1}_{K}[\widehat{H}](z) = \frac{1}{2\pi}\int_{-K}^{K} \left(\frac{1}{ik} + \pi \delta(k)\right)e^{ikz}\di k = \frac{1}{2} + \frac{1}{2\pi}\int_{0}^{K}\left[ \frac{e^{ikz}-e^{-ikz}}{ik} \right] \di k,
\end{equation}
which we may write as
\begin{equation}
    H_{K}(z) = \frac{1}{2} + \frac{1}{\pi}\int_{0}^{K}\frac{\sin(kz)}{k} \di k = \frac{1}{2} + \frac{1}{\pi}\int_{0}^{K z}\frac{\sin(u)}{u} \di u = \frac{1}{2} + \frac{\Si(K z)}{\pi}.
\end{equation}
We formulated a Padé approximation of the $\text{Si}$ function in Maple, following \cite{rowe2015galsim}. We obtain an approximation for $x\in[-2.3,2.3]$ using
\begin{equation}
    \text{Si}(x) = x\cdot\left(\frac{\begin{array}{ll}1&+x^2 \{-4.16045660913051\times10^{-2} \\&+x^2 (9.86950559158117\times10^{-4}\\&+x^2 [-9.98260398342126\times10^{-6}\\&+x^2 \{4.81542491189951\times10^{-8}\\&+x^2(-8.99432372444805\times10^{-11})\}])\}\end{array}}{\begin{array}{rl}1&+x^2 \{1.39509894642504\times10^{-2}\\&+x^2 (9.53388627275856\times10^{-5}\\&+x^2 [4.07021596174597\times10^{-7}\\&+x^2 \{1.11222895369314\times10^{-9}\\&+x^2(1.60360248325839\times10^{-12})\}])\}\end{array}}\right).
\end{equation}
For $x$ values outside that range, we use the relation
\begin{equation}
    \text{Si}(x)=\frac{\pi}{2}-f(x)\cos(x)-g(x)\sin(x),
\end{equation}
with $f(x)=\text{Ci}(x)\sin(x)+[\pi/2-\Si(x)]\cos(x)$ and $g(x)=-\text{Ci}(x)\cos(x)+[\pi/2-\Si(x)]\sin(x)$, where $\text{Ci}$ is the cosine integral relation $\text{Ci}(x)=-\int_{x}^\infty \cos(t)/t \di t$. These auxiliary functions can be found in mathematical reference works. The auxiliary functions can be approximated using a Padé approximation procedure for the region $x'\in[0,1/2.3]$ as an input to $f(1/x')$ and $g(1/x')$. Inverting the obtained expressions in their argument allows us to use the result for $x\in[2.3,\infty]$ in the following approximations
\begin{align}\begin{split}
    f(x)&=\left(\frac{\begin{array}{rl}&8.76577957656212\times 10^{-2}\\
    +&x (3.41402258238831\times 10^{1}\\
    +&x (1.42357662863517\times 10^{2}\\
    +&x (1.59787502038397\times 10^{2}\\
    +&x (7.03579911255664\times 10^{1}\\
    +&x (1.64949384718444\times 10^{1}\\
    +&x (2.28624888100692\times 10^{0}\\
    +&x (1.92842523449289\times 10^{-1}\\
    +&x (9.41205789559801\times 10^{-3}\\
    +&x (2.23396200314283\times 10^{-4})))))))))\end{array}}{\begin{array}{rl}&x( 2.31300105233788\times 10^{1}\\
    +&x (1.35358811016980\times 10^{2}\\
    +&x (2.43359383546479\times 10^{2}\\
    +&x (1.89060299847209\times 10^{2}\\
    +&x (7.47422708097091\times 10^{1}\\
    +&x (1.68761553399714\times 10^{1}\\
    +&x (2.30507299852978\times 10^{0}\\
    +&x (1.93289315843730\times 10^{-1}\\
    +&x (9.41205789560655\times 10^{-3}\\
    +&x (2.23396200314282\times 10^{-4}))))))))))\end{array}}\right)
    \end{split}\end{align}
    and
    \begin{align}\begin{split}
    g(x) &= \left(\frac{\begin{array}{rl}&-8.32280818835294\times 10^{-2}\\
    +&x (2.25386141512083\times 10^{0}\\
    +&x (5.20840820481407\times 10^{1}\\
    +&x (1.09328851692030\times 10^{2}\\
    +&x (4.64218699919365\times 10^{1}\\
    +&x (9.73185708298441\times 10^{0}\\
    +&x (1.19330218679743\times 10^{0}\\
    +&x (8.93684959082754\times 10^{-2}\\
    +&x (3.89884706241243\times 10^{-3}\\
    +&x (8.30826965804889\times 10^{-5})))))))))\end{array}}{\begin{array}{rl}&x^2 (2.51189885193833\times 10^{1}\\
    +&x (1.47667803988686\times 10^{2}\\
    +&x (2.45539153620874\times 10^{2}\\
    +&x (1.60523538558852\times 10^{2}\\
    +&x (5.32542903990403\times 10^{1}\\
    +&x (1.02610878228560\times 10^{1}\\
    +&x (1.21669527870366\times 10^{0}\\
    +&x (8.98669920492969\times 10^{-2}\\
    +&x (3.89884706247557\times 10^{-3}\\
    +&x (8.30826965804714\times 10^{-5}))))))))))\end{array}}\right).\end{split}
\end{align}
We may simply compute the value for negative inputs following $\Si(-x)=-\Si(x)$. The presented approximation for $\Si(x)$ is accurate to within 16 digits, i.e., to within machine precision, for all possible $x$ inputs. The polynomials are rapidly evaluated with modern PC architecture.

\section{Simplified compliance and stiffness matrix relations in isotropic media}\label{app:simplifiedcompliance}
In the main text, we describe how the anti-aliased step-function and 2-D lowpass filters require a filtering of the compliance matrix. Here, we shortly show how these matrices can be computed in a simple  manner for acoustic and elastic isotropic media.
\begin{enumerate}
\item For acoustic media, the compliance matrix is defined as
\begin{align}\label{eq:acousticcompliance}
    \begin{pmatrix} C_{11}(\bx) & C_{13}(\bx) & C_{15}(\bx) \\ C_{13}(\bx) & C_{33}(\bx) & C_{35}(\bx) \\ C_{15}(\bx) & C_{35}(\bx) & C_{55}(\bx) \end{pmatrix}^{+}_\text{acoustic} &= 
    \frac{1}{4}\begin{pmatrix} \kappa(\bx) & \kappa(\bx) & 0 \\ \kappa(\bx) & \kappa(\bx) & 0 \\ 0 & 0 & 0 \end{pmatrix},
\end{align}
which implies that we can anti-alias the fine $\kappa(\bx)$ grid, which we will call $K_a(\bx)$, and take the pseudoinverse of eq. \eqref{eq:acousticcompliance} to reobtain the stiffness matrix as
\begin{align}
    \begin{pmatrix} C_{11}(\bx) & C_{13}(\bx) & C_{15}(\bx) \\ C_{13}(\bx) & C_{33}(\bx) & C_{35}(\bx) \\ C_{15}(\bx) & C_{35}(\bx) & C_{55}(\bx) \end{pmatrix}_\text{acoustic} &= 
    \begin{pmatrix} \frac{1}{K_a(\bx)} & \frac{1}{K_a(\bx)} & 0 \\ \frac{1}{K_a(\bx)} & \frac{1}{K_a(\bx)} & 0 \\ 0 & 0 & 0 \end{pmatrix}.\label{eq:acousticcompliancereinverted}
\end{align}
\cite{mittet2017} obtained this result without resorting to a compliance matrix. The method of the compliance matrix, however, generalizes the findings in \cite{mittet2017} in a simple way.
\item For isotropic media, the compliance matrix is defined as
    \begin{align}\label{eq:isotropiccompliance}
    \begin{pmatrix} C_{11}(\bx) & C_{13}(\bx) & C_{15}(\bx) \\ C_{13}(\bx) & C_{33}(\bx) & C_{35}(\bx) \\ C_{15}(\bx) & C_{35}(\bx) & C_{55}(\bx) \end{pmatrix}^{+}_\text{isotropic} &= 
    \frac{1}{4}\begin{pmatrix} \frac{1}{\lambda(\bx)+\mu(\bx)}+\frac{1}{\mu(\bx)} & \frac{1}{\lambda(\bx)+\mu(\bx)}-\frac{1}{\mu(\bx)} & 0 \\ \frac{1}{\lambda(\bx)+\mu(\bx)}-\frac{1}{\mu(\bx)} & \frac{1}{\lambda(\bx)+\mu(\bx)}+\frac{1}{\mu(\bx)} & 0 \\ 0 & 0 & \frac{4}{\mu(\bx)} \end{pmatrix}.
\end{align}
As in the acoustic case, we can define an anti-aliased grid of $1/(\lambda(\bx)+\mu(\bx))$ which we call $L_a(\bx)$, and an anti-aliased grid of $1/\mu(\bx)$ which we call $M_a(\bx)$. Taking the inverse of eq. \eqref{eq:isotropiccompliance} then reobtains the stiffness matrix as
\begin{align}
        \begin{pmatrix} C_{11}(\bx) & C_{13}(\bx) & C_{15}(\bx) \\ C_{13}(\bx) & C_{33}(\bx) & C_{35}(\bx) \\ C_{15}(\bx) & C_{35}(\bx) & C_{55}(\bx) \end{pmatrix}_\text{isotropic} &= \begin{pmatrix} \frac{1}{L_a(\bx)} + \frac{1}{M_a(\bx)} & \frac{1}{L_a(\bx)} - \frac{1}{M_a(\bx)} & 0 \\ \frac{1}{L_a(\bx)} - \frac{1}{M_a(\bx)} & \frac{1}{L_a(\bx)} + \frac{1}{M_a(\bx)} & 0 \\ 0 & 0 & \frac{1}{M_a(\bx)}  \end{pmatrix}.\label{eq:isotropiccompliancereinverted}
\end{align}
We can use eq. \eqref{eq:isotropiccompliancereinverted} for FD modeling, once we have the anti-aliased $L_a(\bx)$ and $M_a(\bx)$ grids.
\end{enumerate}
The anti-aliased stiffness matrices can thus be computed without matrix inversion routines for the acoustic and elastic isotropic case.

\section{Effect of a smaller stencil size}\label{app:smallerFDstencilresults}
\begin{figure*}
    \centering
    \includegraphics[width=1\textwidth]{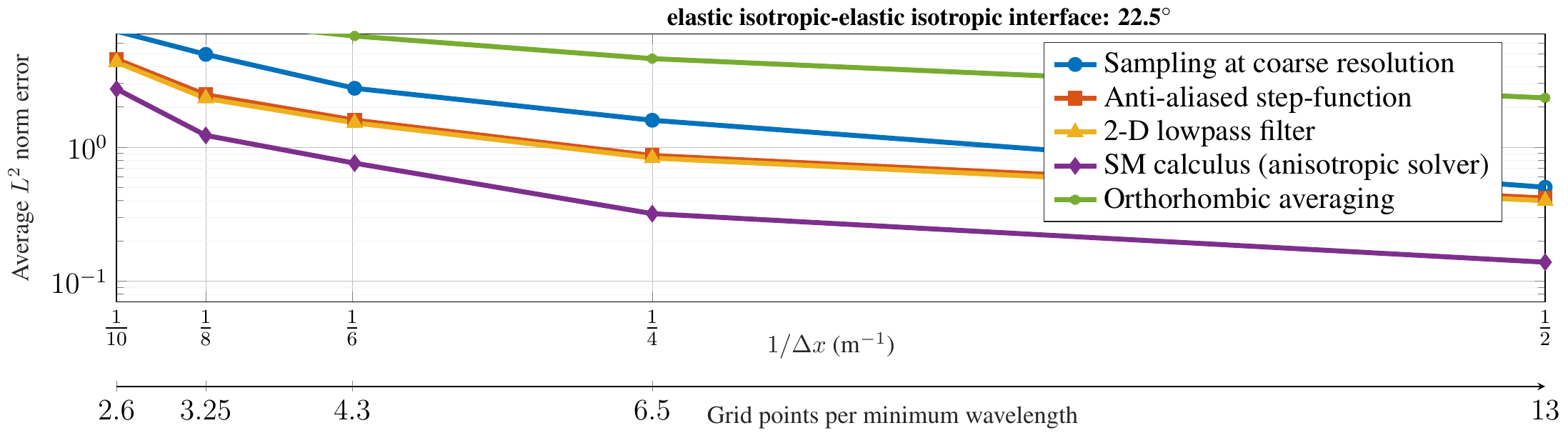}
    \caption{The results for an interface with an angle of $22^\circ$ with respect to the Cartesian grid, for an elastic isotropic-elastic isotropic interface.}
    \label{fig:isotropic_gpw_L3}
\end{figure*}

In the main body of this paper, we used a very large stencil of half-order $L=20$, to ensure that no spatial dispersion errors are introduced by the modeling scheme. Any errors with respect to the analytical solution were then ascribed to the interface representation. In Figure \ref{fig:isotropic_gpw_L3}, we show the results for the elastic isotropic-elastic isotropic interface at $22.5^\circ$, where we use a stencil of half-order $L=3$. The vertical scale is kept identical to Figure \ref{fig:isotropic_gpw}. We may discern that the over-all errors are larger than for the $L=20$ case. However, the relative performance of the five tested methods remains identical to what is found in Figure \ref{fig:isotropic_gpw}. Hence, there is no immediate reason to suspect that the comparison is skewed by the size of the FD stencil.

\end{document}